\documentclass[10pt,twocolumn,twoside,english]{IEEEtran}
\usepackage[T1]{fontenc}
\usepackage{mathrsfs}
\usepackage{bm}
\usepackage{amsmath}
\usepackage{amsthm}
\usepackage{amssymb}
\usepackage{comment}
\usepackage{authblk}
\usepackage{graphicx}
 \PassOptionsToPackage{normalem}{ulem}
\usepackage{bbm}
\makeatletter

\usepackage{tabularx}

\usepackage[switch]{lineno}
\usepackage[named]{algo}
\usepackage{algpseudocode}
\usepackage{algorithm}
\usepackage{bbm}
\usepackage{multirow}

\theoremstyle{plain}
\newtheorem{theorem}{\protect\theoremname}
  \theoremstyle{plain}
  
  \theoremstyle{plain}
  
  \theoremstyle{plain}
   \newtheorem{lemma}{\protect\lemmaname}
  \theoremstyle{remark}
  \newtheorem{remark}{\protect\remarkname}
   
\usepackage{amsmath}
\usepackage{amssymb}
\usepackage{graphicx,psfrag,cite,subfigure}
\usepackage[table]{xcolor}
\usepackage{relsize}
\usepackage[english]{babel}
\usepackage[autostyle]{csquotes}
\usepackage{soul}

\setkeys{Gin}{width=1.0\columnwidth}

\makeatother
  \providecommand{\definitionname}{Definition}
  \providecommand{\lemmaname}{Lemma}
  \providecommand{\propositionname}{Proposition}
  \providecommand{\remarkname}{Remark}
\providecommand{\theoremname}{Theorem}
\providecommand{\conjecturename}{Conjecture}
\providecommand{\assumptionname}{Assumption}

\begin{document}
%
\title{Design and Detection of   
Controller Manipulation Attack on RIS  Assisted Communication}
%
%
%


\author{ Siddharth~Sankar~Acharjee \& Arpan~Chattopadhyay
\thanks{S.S.A. is currently working in Mediatek Bangalore. This work was done when S.S.A was with Bharti School of Telecommunications Technology and Management (BSTTM), Indian Institute of Technology (IIT), Delhi. Email: sidacharjee95@gmail.com.\\
A.C.  is with the Department of Electrical Engineering and BSTTM, IIT Delhi. Email: arpanc@ee.iitd.ac.in}
\thanks{A. C. acknowledges support via  grant no. GP/2021/ISSC/022 from IHFC, Delhi and grant no. CRG/2022/003707 from SERB, India.}
}

\maketitle
\pagestyle{empty}
\thispagestyle{empty}
\begin{abstract}
In recent years, research on signal and information theory from the electromagnetics viewpoint has drawn  significant attention, mostly due to its potential use in various communication technologies such as multiple-input-multiple-output (MIMO) and Reconfigurable Intelligent Surface (RIS).  In this paper, we introduce a new attack called controller manipulation attack (CMA) on a  RIS  assisted communication system between a transmitter and a receiver, and develop mathematical theory for its design and detection. An attacker has the capability to manipulate the RIS controller and modify the phase shift induced by the RIS elements on the incident electromagnetic signal. The goal of the attacker is to minimize the data rate at the receiver, subject to a constraint on the attack detection probability at the receiver. We consider a number of   attack detection models: (i) composite hypothesis testing based attack detection in a given  fading block for known channel gains, (ii)   quickest detection of CMA in a given  fading block for known channel gains, (iii) nonparametric hypothesis test to detect CMA for unknown channel gains over a fading block, and (iv) signal-to-noise-ratio (SNR) moment based detection over possibly multiple fading blocks. In the first case, we show that a simple energy detector is uniformly most powerful (UMP).  In the second case, simplification of the standard CUSUM test and its performance bounds are obtained. In the third case, non-parametric Kolmogorov-Smirnov test is further simplified to a simple per-sample double threshold test. The optimal attack against these three    detectors are designed via   novel optimization formulations and   semidefinite relaxation based solutions. In the fourth case, we consider threshold  detection using  moments of SNR; various SNR moments under no attack are obtained analytically for large RIS and then used to formulate the optimal attack design problem as a linear program.  Finally, numerical results demonstrate the efficacy of the proposed schemes.
\end{abstract}

%
\IEEEpeerreviewmaketitle

\section{Introduction}\label{Introduction}
\begin{figure}[t!]
\begin{centering}
\begin{center}
\includegraphics[height=4.5cm, width=0.90\linewidth]{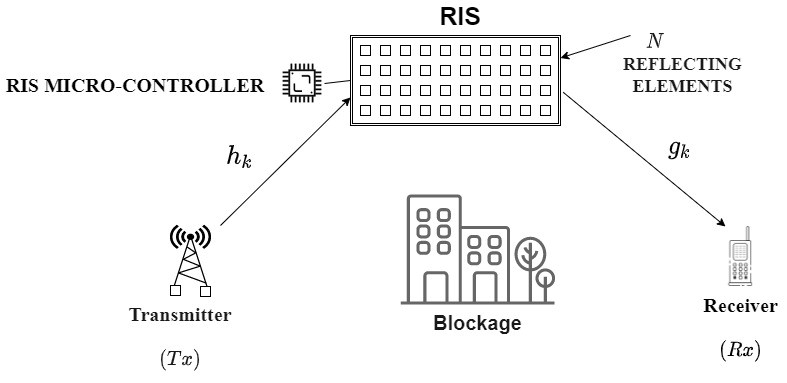}
\end{center}
\end{centering}
\caption{RIS assisted wireless communication model.}
\label{fig:Single antenna model}
\end{figure}

%
%
%
%
\IEEEPARstart{T}{he} last two decades have seen tremendous activities towards developing techniques that exploit the inherent randomness of the wireless 
propagation medium. Such research efforts  traditionally focused on optimizing the transmission and reception schemes.  However, reconfigurable metasurfaces have shown the potential to be useful as a  tool for controlling the wireless propagation medium. Certain electromagnetic (EM) properties of these surfaces can be electronically 
controlled, which helps in their use in passive beamforming for wireless communication without using additional transmit antennas and radio frequency (RF) chains.  Some well-known implementations of RIS  include reflect arrays and software defined metasurfaces \cite{di2020smart,docomo2020docomo,arun2020rfocus,liu2021reconfigurable,bjornson2020reconfigurable}, 
and this technology is envisioned to play a key role in 6G communications. The RIS consists of multiple reflecting elements,  and the phase shift  on the incident  EM wave, induced by reflection  at each   element, can be electronically controlled by an RIS micro-controller;  see Figure~\ref{fig:Single antenna model}. By  simultaneously adjusting the phase shifts of all elements, the RIS can fully control the strength and direction of the reflected EM waves. A distinct advantage of RISs over
the existing technologies is that, RIS enables the network administrator to control the propagation
environment to a certain extent, thereby leading to a boost in
signal strength at the receiver. Some other key features which distinguish RIS  from other technologies are as follows: (i) RIS does not need a dedicated power source assigned to it and is almost passive in nature, (ii) RIS is immune to receiver noise as it does not require up/down conversions or complex power amplifiers, and hence neither amplifies   nor contributes any noise to the reflected signals, (iii) RIS can be easily installed on buildings, billboards, indoor spaces etc. However, the above features also give rise to  many design and implementation related
challenges and prospects \cite{zheng2022survey}. For example, active transmit beamforming at the transmitter and passive reflect beamforming at the RIS can be used to optimize performance metrics such as transmit power   and energy efficiency \cite{wu2019intelligent,wu2019beamforming}.
Lately, the need for high bit rate communication at higher frequency bands such as terahertz (THz)  has given rise to many additional  challenges. Smaller wavelength leads to higher susceptibility  to blockages and scattering effects, and, combined with larger antenna or RIS aperture, results in a large  Fraunhofer distance. This necessitates communication in the near-field regime involving spherical  propagation model for the EM waves, in contrast to planar wave model in far-field communication, and gives rise to a number of design and implementation related problems \cite{cui2022near}. From the viewpoint of EM theory, \cite{tang2020wireless} considers a system model for RIS-assisted wireless communication that takes into consideration many physical parameters such as  physical size  and  radiation pattern of the RIS,   develops a free space path loss model for the RIS aided communication system in near and far field regimes, and validates it experimentally. By using the model in  \cite{tang2020wireless}, the paper  \cite{cheng2021joint}  optimizes system performance over  transmit beamforming, RIS phase shift  and RIS positioning. The need for a better understanding of RIS and MIMO communication systems in the near field regime for higher frequency bands resulted in a number of works investigating the connection between electromagnetic signal and information theory \cite{wan2021mutual, zhu2022electromagnetic, migliore2008electromagnetics}.
 Also, there has been a significant number of works on holographic RIS and its use in  THz  communication. Holographic RIS consists  of a large number of tiny and inexpensive antennas or reconfigurable elements that are densely packed in a small space in order to realize a holographic array with a spatially continuous aperture; see  \cite{wan2021terahertz} for derivation of the optimal beam pattern and RIS  reflection coefficients in this context.  The effect of electromagnetic interference (EMI) in  RIS aided communication has been studied in \cite{de2021electromagnetic}. 

Meanwhile, physical layer security has been intensively explored  as a counterpart to upper layer encryption approaches towards   wireless security. A significant number of such works focus on enhancing the secrecy performance  against an eavesdropper; e.g., \cite{guan2020intelligent} that combines  transmit beamforming  with artificial noise in an RIS based setup,  \cite{yu2020robust}   for using  RIS to maximise the secrecy rate, and \cite{mishra2022optm3sec} to optimize secrecy rate in a dual function radar and communication system assisted by RIS. An eavesdropper can even place a  metasurface   between a transmitter and the receiver, in order to steer the electromagnetic wave towards itself  \cite{shaikhanov2022adversarial}.   The scenario where an eavesdropper can use EMI to degrade the secrecy performance at the legitimate user, and  increase the signal leakage to itself, is studied in \cite{vega2022physical}. Additionally, in certain defence applications, even the transmission of data can raise suspicion; this leads to the need for covert communication. A significant number of works are dedicated to ensure  covert communication between a transmitter (Alice) to a receiver  (Bob) while keeping the communication undetected to an Warden (Willie) \cite{bash2013limits,bloch2016covert}. 
Wireless transmissions are also subject to jamming attacks\cite{cheng2017time,osanaiye2018statistical}; a traditional jammer uses its own  energy source to broadcast powerful noise signals to the victim node. Interestingly,  the authors of \cite{lyu2020irs} considered  an RIS based jammer. Another  attack is pilot contamination attack (PCA), where the pilot sequence is manipulated by an eavesdropper using RIS during the channel estimation phase, leading to incorrect channel state information (CSI) at the transmitter 
and consequently significant signal leakage to the eavesdropper during the succeeding data transmission phase; see  \cite{huang2020intelligent}. On the other hand, false data injection (FDI) attack \cite{chattopadhyay2019security, choraria2022design} is a   threat to networked control systems, where the data flowing in the system is manipulated by an external    attacker to disrupt the estimation and control operations.

In this paper, we propose a new physical layer attack called controller manipulation attack (CMA) on RIS-assisted far field communication, where an external attacker can manipulate the phase shift induced by the RIS during the data transmission phase. In spirit,  CMA is a combination of PCA and FDI. However, unlike PCA, CMA works on the data  and not on the pilot symbols, and unlike FDI, CMA operates at the physical layer. CMA can be carried by an attacker that can either hack the RIS microcontroller or infect it by a malware, or physically replace it. As seen later in the paper, CMA results in significant throughput loss to the communication system. The goal of this paper is to develop a mathematical theory for CMA detection and design. While there has been a number of works investigating the connection between electromagnetic signal and information theory, our current work takes a different route and explores information rate manipulation by an attacker via controlling the electromagnetic signal propagation environment.  
We summarise our major contributions as follows:
\begin{enumerate}
    \item We propose the CMA attack for the first time, and provide its mathematical model for attack detection and optimal attack design.
    \item In Section~\ref{Attack design against a detector using UMP test}, for the case where   multiple received data symbols in a single fading block are used to detect CMA, we show that a simple energy detector is uniformly most powerful (UMP). In the process, we discover a certain interesting relation between the received signal variance, false alarm probability and attack detection probability. Next, the energy detector is leveraged to formulate an optimization problem for attack design to minimize the received SNR  subject to attack detection constraint. This non-convex problem is converted to a rank-constrained optimization problem, which is sub-optimally solved via a   combination of semidefinite relaxation and Gaussian randomization.
    \item Next, in Section~\ref{Sequential Detection}, we consider the scenario where the data symbols are received sequentially during a particular fading block, and develop the quickest attack detection strategy using the  Generalised Likelihood Ratio-Cumulative Sum  (GLR-CUSUM) algorithm. We derive bounds for average run length to false alarm (ARLFA) and the average run length to detection (ARLD), which are later used in identifying tunable hyper-parameters for attack design. We also show that   attack can be designed by solving a   similar optimization problem to that used for fixed sample size  in Section~\ref{Attack design against a detector using UMP test}.
    \item In Section~\ref{joint design}, we show that joint beamforming by a multi-antenna transmitter and an RIS can be handled in a similar way as the single antenna case. The detector's structure remains similar, and the attack can be
designed by solving an optimization problem very similar to that   in Section~\ref{Attack design against a detector using UMP test}.
    \item In Section~\ref{Attack at a large time window}, we develop the theory of CMA for detection over  multiple fading blocks, when channel gains are known only to the attacker. The detector at the receiver employs a threshold based detection rule   on the difference between the empirical SNR moments and the SNR moments under no attack. The attack design  problem is shown to be a linear program. In the process, we derive closed form expressions of the moment generating function  and the first and second order moments of the received SNR under no attack.
    \item In Section~\ref{CSI uncertainty}, we derive the distribution of the received signal as well as its energy under imperfect CSI  in closed form, which are  then used to design a  non-parametric test based detector. The detector in this case can be simplified to obtain a per-sample double  threshold test which selects the alternate hypothesis for attack if the received signal sample energy is either below a lower threshold or above an upper threshold, which is markedly different from the usual energy detector. We also show that attack design against this detector involves solving a similar optimization problem as before. Finally, we discuss the possible extensions and challenges for non-stationary channel under imperfect CSI and also for quickest detection of CMA over a fading block under imperfect CSI.
\end{enumerate}
After attack detection, the communication system can either   install an anti malware program in the RIS microcontroller, or disconnect the power supply to the compromised RIS. If there are multiple RIS units, safe RIS units can be instructed to adjust their beamforming algorithms to compensate for the compromised RIS. However, CMA in the multi-RIS setting is left for future research. Depending on the topography of the deployment region, the attacker may or may not be able to see other reflectors. However, the CMA attacks are electronic in nature, and is typically performed by a remote agent by hacking the controller of the victim RIS. The remote attacker may not be able to see other RIS units in such cases. Even if the attacker can see the other RIS, it will not be able to control them unless it hacks their controllers. Also, the attack-defence dynamics unfold at a much faster timescale compared to the speed at which manual intervention functions.
The rest of the paper is organized as follows. The system model is described in Section~\ref{System Model}. Theoretical analyses under various modeling assumptions are provided between Sections~\ref{Attack design against a detector using UMP test}-\ref{CSI uncertainty}. Numerical results are provided in Section~\ref{NumericalResults}, followed by the conclusions in Section~\ref{Conclusion} and  proofs  in the appendices.

A preliminary version of this paper appeared in \cite{acharjee2022controller} which dealt with    attack detection and design during a given fading block  for a fixed sample size based detector, and also attack over multiple fading blocks. However, no mathematical proof was provided in \cite{acharjee2022controller}. In this paper, we additionally develop theory for a sequential detector performing quickest detection of CMA in  Section~\ref{Sequential Detection}, attack design against joint beamforming by   transmitter and RIS in Section~\ref{joint design}, and attack detection and design under imperfect CSI  in Section~\ref{CSI uncertainty}. The proofs of all mathematical results including those of \cite{acharjee2022controller} are also new contributions.

\textit{Notation}: In this work, bold capital letters, bold lowercase letters, and calligraphic letters will represent matrices, vectors, and sets respectively. Transpose and conjugate transpose are denoted by $(\cdot)^{\mathsmaller T}$ and $(\cdot)^{\mathsmaller H}$. The notation  $\operatorname{diag}(\mathbf{x})$ represents a diagonal matrix in which each diagonal element corresponds to an element in $\mathbf{x}$, whereas   $||\mathbf{x}||$ denotes the norm of $\mathbf{x}$. The expectation, probability, identity matrix and zero matrix are represented by $\mathbb{E}$, $\mathbb{P}$, $\bm{I}$ and $\bm{0}$ respectively. Supremum and the essential supremum are represented by $\sup$ and $\operatorname{ess} \sup$,  respectively.
\section{System Model}\label{System Model}
We consider a wireless communication system where  a single antenna transmitter (\textit{Tx}) communicates with a single antenna receiver (\textit{Rx}) and is assisted by an RIS consisting of $N$ elements. However, extension to   multiple transmit antennas is considered in Section~\ref{joint design}. A flat fading model is assumed for all the channels involved. Furthermore, we assume that the line-of-sight (LOS) path is not present between the transmitter and receiver due to severe blockage,  and the signals that are reflected from the RIS more than once have negligible strength \cite{wu2019intelligent}. We denote the channel between the transmitter and the $k^{\text{th}}$ RIS element by $h_k$,  whereas the channel between the $k^{\text{th}}$ RIS element and the receiver is denoted by $g_k$.
We consider independent Rayleigh fading along the lines of \cite{9404225}, i.e.,  
$h_{k} = \sqrt{\epsilon_{h}} \tilde{h}_{k}$ and
$g_{k} = \sqrt{\epsilon_{g}} \tilde{g}_{k}$, where   $\tilde{h}_{k}
\sim \mathcal{CN}(0,1)$, $\tilde{g}_{k}
\sim \mathcal{CN}(0,1)$, and  $\epsilon_{h}$, $\epsilon_{g}$ represent the corresponding path losses. Let $
h_{k}\doteq \alpha_{k} e^{j \theta_{k}}$, $g_{k} \doteq \beta_{k} e^{j \psi_{k}}$   where $\alpha_{k} =|h_{k}| $ and $\beta_{k} =|g_{k}|$ represent the channel amplitudes following Rayleigh distribution, while $\theta_{k}$ and $\psi_{k}$ are the equivalent channel phases, uniformly distributed between $[0, 2\pi ]$.

Let the phase shift induced by the $k^{\text{th}}$ RIS element be $\phi_k$. The effect of the RIS can be captured by a matrix $\boldsymbol{\Phi}=\operatorname{diag}([\mathrm{e}^{j \phi_{1}}, \ldots, \mathrm{e}^{j \phi_{N}}])$, and let  $\bm{\Omega} \doteq [  \phi_{1},\phi_{2} \ldots, \phi_{N}]^{T}$ be a vector consisting of the corresponding RIS phase shifts. The baseband signal at the receiver is given by (see \cite{kudathanthirige2020performance}):
\begin{equation}\label{receivedsignal}
y=\sqrt{P} \sum_{k=1}^{N} g_{k}\mathrm{e}^{j \phi_{k}} h_{k} x+w
\end{equation}
Here ${x}\sim \mathcal{CN}(0,1)$ represents the transmitted signal drawn from a Gaussian codebook, $P$ denotes the transmit power, and   ${w}\sim \mathcal{CN}(0,\sigma_{w}^2)$ denotes the additive white Gaussian noise (AWGN) at the receiver. We can equivalently write \eqref{receivedsignal} as (see \cite{basar2019wireless}): 
\begin{equation}\label{receivedsignalvectorrep}
y=\sqrt{P}\mathbf{g}^{{\mathsmaller H}} \mathbf{\Phi} \mathbf{h} x+w
\end{equation}
\noindent where $\mathbf{h} \doteq \left[\bar{h}_{1}, \bar{h}_{2} \ldots \bar{h}_{N}\right]^{\mathsmaller H}$, where $\bar{h}_{k}$ denotes the complex conjugate of $h_k$, and $\mathbf{g} \doteq \left[g_{1}, g_{2} \ldots g_{N}\right]^{\mathsmaller H}$. Let $\Lambda_k \doteq \phi_{k}+\psi_{k}+\theta_{k}$.
The received SNR can be expressed as $\Gamma=\bar{\kappa}\left|\mathbf{g}^{\mathsmaller H} \mathbf{\Phi} \mathbf{h}\right|^{2}$
where $\bar{\kappa} \doteq P/{\sigma_{w}^2}$, which can further be rewritten as:
\begin{equation}\label{receivedsnr2}
\Gamma=\bar{\kappa}\left|\sum_{k=1}^{N} \alpha_{k}  \beta_{k} \exp \left(j\left[\phi_{k}+\psi_{k}+\theta_{k}\right]\right)\right|^{2}
\end{equation}
We consider two phase shift models:
      \par \textit{Continuous phase shift model:} Here, the RIS elements are able to generate any arbitrary phase shift between $[0, 2\pi ]$. If the phase shifts generated by the combined propagation channel through the RIS is accurately known,  the reflection phase shift of each element is adjusted by the RIS microcontroller to achieve zero phase error. Obviously, when the RIS sets its phases   as $\phi_{k}^{*}=-\left(\theta_{k}+\psi_{k}\right)$  optimum SNR is achieved at the receiver. We denote ${\bm{\Phi}_{0}} \doteq \operatorname{diag}([\mathrm{e}^{j \phi_{1}^{*}}, \ldots, \mathrm{e}^{j \phi_{N}^{*}}])$ as a matrix corresponding to the optimum phase shifts at the RIS.
        \par \textit{Discrete phase shift model:} Here the RIS elements are able to realise only a finite, discrete set of phase shifts. Typically, for each element, the collection of possible phase shifts is given by $\mathcal{D}=\{0,\Delta \phi,\ldots,\Delta\phi(M-1)\}$, where $\Delta\phi=2\pi/M$ and $M=2^b$ for   $b \in \mathbb{Z}_+$; see  \cite{wu2019beamforming}, \cite{9404225}. Let $\hat{\phi}_k \in \mathcal{D}$ be the phase shift induced by the $k^{\text{th}}$ element of the RIS. We define the quantization error for the $k^{\text{th}}$ element as $\delta_{k}=\hat{\phi_{k}}-{\phi_{k}}^{*}$, where $\phi_{k}^{*}$ is the optimum phase shift under the continuous phase shift model. Since   Rayleigh fading channel gains $g_k$ and $h_k$ are independent across $k$, and the elements in $\mathcal{D}$ are spaced in regular intervals,   $\delta_{k}$ is independently and uniformly distributed over $(-\tau, \tau]$,  $\tau=\pi/2^b$ \cite{haykin2008communication}.\\
\vspace{-0.5mm}
{\bf Modeling differences and analytical commonalities among various   sections:} It is to be noted that Sections~\ref{Attack design against a detector using UMP test}-\ref{CSI uncertainty} make different modeling assumptions, with Section~\ref{Attack design against a detector using UMP test} assuming the most basic model.  While Section~\ref{Sequential Detection} deals with quickest detection of CMA, other sections primarily deal with fixed sample size detectors. However, 
we broadly consider two scenarios for the attacker who adversarially controls $\bm{\Omega}$ in order to hamper the communication.  In the first scenario, attack  detection  is carried out by the receiver in each fading block  at  a much faster   timescale, and, taking this into account, the attacker seeks to minimise the data rate. In this scenario, Section~\ref{Attack design against a detector using UMP test} considers the most elementary setup with fixed sample size detector operating over a fading block, which is later 
extended to multiple transmit antennas (Section~\ref{joint design}) and imperfect CSI (Section~\ref{CSI uncertainty}).   On the other hand, in  Section~\ref{Sequential Detection}, a sequential detector is considered for quickest detection over a given fading block.  Here the channel gains are assumed to be known to the attacker and the detector.   In the second scenario, attack detection  is carried out by the receiver over multiple fading blocks or at a much slower timescale,  and  the attacker seeks to minimise the ergodic data rate at the receiver; see  Section~\ref{Attack at a large time window}.  In each of these sections, we first propose a reasonable attack detector, analyse its performance and optimality,  formulate the  attack design problem against this detector as an optimization problem,  and provide an algorithm to solve it. It turns out that the optimization problems in Sections~\ref{Attack design against a detector using UMP test}, \ref{Sequential Detection}, \ref{joint design} and \ref{CSI uncertainty} for attack design are very similar, but it is different for Section~\ref{Attack at a large time window}. 

\section{Attack for a given fading block, fixed sample size detector, perfect CSI}\label{Attack design against a detector using UMP test}
In this section, we assume that the receiver and the attacker have perfect knowledge of the channel gains $\{h_k,g_k\}_{1 \leq k \leq N}$ over a quasi-static flat fading block under consideration, and  $P=1$. The transmitted symbols are drawn from a Gaussian code-book, and  hence ${x}
\sim \mathcal{CN}(0,1)$. Obviously, the conditional distribution of $y$ given $\mathbf{h}$ and $\mathbf{g}$,  can be written as  ${y}
\sim \mathcal{CN}(0,\sigma_w^2+\left|\mathbf{g}^{\mathsmaller H} \mathbf{\Phi} \mathbf{h}\right|^{2})$. We define $\sigma^2 \doteq \sigma_w^2+ \left|\mathbf{g}^{\mathsmaller H} \mathbf{\Phi} \mathbf{h}\right|^2 $ and  $\sigma_0^2 \doteq \sigma_w^2+ \left|\mathbf{g}^{\mathsmaller H} {\bm{\Phi}_{0}} \mathbf{h}\right|^2$ as the received signal variances under attack and under  no attack, respectively. We consider continuous phase shift model in this section. 

\begin{remark}
We have considered $P=1$ for  mathematical convenience, but the  analysis in this paper will hold for $P \neq 1$. In that case, we will have $\sigma^2 \doteq \sigma_w^2+ P \left|\mathbf{g}^{\mathsmaller H} \mathbf{\Phi} \mathbf{h}\right|^2 $ and  $\sigma_0^2 \doteq \sigma_w^2+ P \left|\mathbf{g}^{\mathsmaller H} {\bm{\Phi}_{0}} \mathbf{h}\right|^2$.
\end{remark}
\begin{lemma}\label{lemma:maxvariance}
$\sigma^2 \leq \sigma_{0}^2$ if $\bm{\Phi} \neq \bm{\Phi}_{0}$.  Equality holds  if $\bm{\Phi} = \bm{\Phi}_{0}$. \qed
\end{lemma}

Lemma \ref{lemma:maxvariance} provides  an intuition that any non-optimal phase shift at the RIS   results in a smaller received  signal variance than the optimal case; this observation crucially helps in designing the CMA detector at the receiver. However, it has to be noted that $\{{\phi_k}^{*}+c\}_{1 \leq k \leq N}$ for any constant $c$ also maximizes the variance of $y$, and hence we consider solutions with $\Lambda_k=0 \,\, \forall 1 \leq k \leq N$ without loss of generality.

\subsection{Attack detection via hypothesis testing}\label{Attack Detection UMP}
We formulate the detection problem   as a binary hypothesis testing problem. Let us assume that $K$ symbols are transmitted over a fading block, and the receiver detects a possible CMA at the end of the fading block. The two hypotheses are:
\begin{align*}
 H_0:  \bm{\Phi}=\bm{\Phi}_0 \hspace{8mm} \text{and} \hspace{8mm} H_1:  \bm{\Phi} \neq \bm{\Phi}_0
\end{align*}
Clearly, the null hypothesis means that there is no attack, and the alternate hypothesis implies an attack. Since the receiver receives $K$ i.i.d observations $\{{y}_i\}_{1 \leq i \leq K}$ in a fading block, where $y_i \sim \mathcal{CN}(0, \sigma^2)$, the two hypotheses can be equivalently represented as:
\begin{align*}
H_0: \sigma=\sigma_0 \hspace{8mm} \text{and} \hspace{8mm} H_1:  \sigma < \sigma_0
\end{align*}
\begin{lemma}\label{lemma:likelihoodratio}
The optimum likelihood ratio test reduces to 
\begin{equation}
    W \doteq \sum_{i=1}^{K}||y_i||^2 
    {\underset{H_{0}}{\overset{H_{1}}{\lesseqqgtr}}} \eta'
\end{equation}
where $\eta'>0$ is a suitable threshold. \qed
\end{lemma}

In Lemma \ref{lemma:likelihoodratio}, we observe that the likelihood ratio test simplifies to an energy detector; when the received signal power (or SNR) is low, the detector declares an attack.  
\begin{lemma}
\label{lemma:Yidistribuition}
If $y_{i}\sim \mathcal{CN}(0,{\Tilde{\sigma}}^2)$ for any $\Tilde{\sigma} > 0$, then   $2\lambda W$ follows the Chi-squared distribution ($\chi^2$ distribution) with $2K$ degrees of freedom, where $\lambda=\frac{1}{\Tilde{\sigma}^2}$ and $W \doteq \sum_{i=1}^{K}||y_i||^2 $. \qed
\end{lemma}
The probability of false alarm (PFA) is defined as $P_{FA}=\mathbb{P}(W  \leq \eta' |H_{0})$. Lemma \ref{lemma:Yidistribuition} helps us in deriving $P_{FA}$ of the detector and thus helps in attack design.
Let $\rho$ represent the significance level of a test and $R_{2K,\rho}$ denote the inverse cumulative distribution function of the chi-squared probability density function (PDF) with $2K$ degrees of freedom, evaluated at a probability  $\rho \in [0,1]$.
\begin{theorem}\label{theorem:UMP-Test}
The likelihood ratio test described in Lemma \ref{lemma:likelihoodratio} is a  Universally Most Powerful (UMP \cite[Page~35]{poor1998introduction}  ) test with a threshold  $\eta'={R_{2K,\rho}}{\sigma_{0}^2}$. \qed
\end{theorem}
\subsection{Designing the attack against the UMP detector}\label{Designing the attack against the UMP detector}
Now we  design an attack strategy against the UMP test, assuming that the attacker has knowledge of the threshold $\eta'$. 
The probability of detection ($P_{D}$) can be written as follows:
\begin{equation}\label{PD}
    P_{D}= \mathbb{P}( W \leq \eta'| H_1)  
\end{equation}
From the perspective of an attacker, we seek to find an optimum $\sigma^2$ which is realizable via RIS phase shift, minimizes the data rate at the receiver, and achieves a low detection probability.
\begin{theorem}\label{theorem:Attack Design against UMP}
The value of $\sigma^2$ for a particular probability of detection $\xi$ and a known threshold determined by $\rho$ as per Theorem~\ref{theorem:UMP-Test} is given by  $\sigma^2=\frac{R_{2K,\rho} \sigma_0^2}{R_{2K,\xi}}$. \qed
\end{theorem}
Theorem \ref{theorem:Attack Design against UMP} helps us in finding $\sigma^2$ for a given detection probability $\xi$. Also, the detection probability  decreases as $\sigma^2$ increases (i.e., as $|\sigma_0^2-\sigma^2|$ decreases), since it becomes more difficult to distinguish between the two hypotheses.   Theorem \ref{theorem:Attack Design against UMP} can be used by an attacker to calculate the phase shift vector $\bm{\Omega}$. However, any arbitrarily small $\sigma^2$ may not be realizable by $\bm{\Omega}$.   Hence, given that  $SNR=\frac{|\bm{{g}}^{H}{\bm{\Phi}\bm{h}}|^{2}}{\sigma_w^2}$,  the optimization problem from the attacker's perspective is:
\begin{equation}\label{attackUMPtest}
\begin{aligned}
\min_{\bm{\Omega}} \quad & \log(1+SNR) \hspace{0.4cm}
\textrm{s.t.} \quad P_{D}\leq\xi \hspace{0.3cm}\\
\end{aligned}
\end{equation}
The attacker seeks to minimize an  upper bound to the channel capacity. While  $\log(1+SNR)$ is the channel capacity under   perfect CSI at the transmitter and receiver, CMA will make the overall channel unknown to both of them. However, $\log(1+SNR)$ will still be an upper bound to the data rate.  

\subsubsection{\bf {Procedure to find the attack phase shift vector}}\label{Procedure to find the attack phase shift vector} Since  $P_{D}\leq\xi$ can be simplified using Theorem \ref{theorem:Attack Design against UMP} as $\sigma^2 \geq \frac{{R_{2K,\rho}}{\sigma_{0}^2}}{R_{2K,\xi}} $ or  $|\bm{{g}}^{H}{\bm{\Phi}\bm{h}}|^{2} \geq \frac{{R_{2K,\rho}}{\sigma_{0}^2}}{{R_{2K,\xi}}}-\sigma_w^2 \doteq \nu$, and since $\log(\cdot)$ is an increasing function, we can alternatively write \eqref{attackUMPtest} as follows:
\begin{equation}\label{attackUMPtest_alternate_vectorform}
\begin{aligned}
\min_{\bm{\Omega}} \quad & |\bm{{g}}^{H}\bm{{\Phi}{h}}|^{2}\\
\textrm{s.t.} \quad |\bm{{g}}^{H}\bm{{\Phi}{h}}|^{2} \geq \nu; \hspace{0.1cm}0&\leq \hspace{0.1cm} \phi_{k}\leq 2\pi, \hspace{0.1cm}1 \leq k \leq N
\end{aligned}
\end{equation}
Let $\bm{s}=[s_{1}, s_{2} \ldots s_{N}]^{\mathsmaller H}$ where $s_{k}=\mathrm{e}^{j \phi_{k}} $. We define 
$\bm{\Psi} \doteq \operatorname{diag}(\bm{g}^{H}) \bm{h}$. Hence, we have
\begin{equation}\label{Composite channel}
    \bm{{g}}^{H}\bm{{\Phi}{h}}=\bm{s}^{H}\bm{\Psi}; \hspace{2.5mm}|\bm{{g}}^{H}\bm{{\Phi}{h}}|^{2}=\bm{s}^{H}\bm{\Psi}\bm{\Psi}^{H}\bm{s}
\end{equation}
Thus, \eqref{attackUMPtest_alternate_vectorform} can be written as:
\begin{equation}\label{transformation_to_SDP_form1}
\begin{aligned}
\min_{\bm{s}} \quad & \bm{s}^{H}\bm{\Psi}\bm{\Psi}^{H}\bm{s}\\
\textrm{s.t.} \quad |s_{k}|=1,\hspace{0.1cm}1 \leq &k \leq N;\hspace{0.2cm} \bm{s}^{H}\bm{\Psi}\bm{\Psi}^{H}\bm{s}\geq \nu
\end{aligned}
\end{equation}
We observe that \eqref{transformation_to_SDP_form1} is a non-convex optimization problem, hence we use an auxiliary matrix $\bm{L}=\begin{bmatrix}
\bm{\Psi}\bm{\Psi}^{H} & \bm{0}\\
\bm{0} & 0
\end{bmatrix}$ and an auxiliary vector $\tilde{\bm{s}}=\begin{bmatrix}
\bm{s} \\
1 
\end{bmatrix}$ to tackle this issue. We can further write  ${\tilde{\bm{s}}}^{H}\bm{L}{\tilde{\bm{s}}}=\operatorname{Tr}(\bm{L}{\tilde{\bm{s}}}{\tilde{\bm{s}}}^{\mathsmaller H})=\operatorname{Tr}(\bm{LS})$, where $\bm{S}={\tilde{\bm{s}}}{\tilde{\bm{s}}}^{H}$ and needs to fulfill $\bm{S}\succeq0$ and $\operatorname{rank}(\bm{S})=1$. We can reformulate \eqref{transformation_to_SDP_form1} as follows:
\begin{equation}\label{transformation_to_SDP_form2}
\begin{aligned}
\min_{\bm{S}} \quad & \operatorname{Tr}(\bm{LS})\\
\textrm{s.t.}\hspace{0.4cm} \bm{S}_{k,k}=1,\hspace{0.3cm}&1 \leq k \leq N+1 \\
\bm{S}\succeq0,\hspace{0.15cm} &\operatorname{rank}(\bm{S})=1,\hspace{0.15cm} \operatorname{Tr}(\bm{LS})\geq \nu 
\end{aligned}
\end{equation}
We note that in \eqref{transformation_to_SDP_form2}, the rank one constraint makes the problem non-convex. We apply semidefinite relaxation (SDR) to relax the rank-one constraint:
\begin{equation}\label{transformation_to_SDP_form_final}
\begin{aligned}
\min_{\bm{S}} \quad & \operatorname{Tr}(\bm{LS})\\
\textrm{s.t.} \hspace{0.2cm} \bm{S}_{k,k}=1,\hspace{0.5mm}1\leq k&\leq N+1; \hspace{0.1cm}\bm{S}\succeq0,\hspace{0.1cm} \operatorname{Tr}(\bm{LS})\geq\nu
\end{aligned}
\end{equation}
Clearly, \eqref{transformation_to_SDP_form_final} is  a convex semidefinite program (SDP \cite{boyd2004convex})  and hence can be solved using solvers such as CVX \cite{grant2014cvx}. In fact, for a given accuracy level $\hat{\epsilon}$, solving  \eqref{transformation_to_SDP_form_final} via SDP involves a worst case complexity of  
$\mathcal{O}\left(\max \{1, N\}^4 N^{1/2} \log (1/\hat{\epsilon})\right)$; see   \cite{luo2010semidefinite}.  However, the solution obtained is not necessarily rank-one and hence yields a lower-bound to the objective in \eqref{transformation_to_SDP_form2}. Hence, we proceed to create a rank-one solution by following the process of Gaussian randomization \cite{wu2019intelligent}, \cite{lyu2020irs}. We can write the eigenvalue decomposition of $\bm{S}$ as $\bm{S}=\bm{Q}\bm{\Sigma}\bm{Q}^{H}$ where $\bm{Q}=[\bm{q}_{1},\bm{q}_{2},\ldots,\bm{q}_{N+1}]$ is a matrix consisting of eigenvectors of $\bm{S}$ and $\bm{\Sigma}=\operatorname{diag}(\bar{\lambda}_{1},\bar{\lambda}_{2},\ldots,\bar{\lambda}_{N+1})$ is a diagonal matrix  consisting of corresponding eigenvalues. We can write a sub-optimal solution to \eqref{transformation_to_SDP_form2} as $\tilde{\bm{s}}=\bm{Q}\sqrt{\bm{\Sigma}}\bm{\bar{f}}$, here $\bm{\bar{f}}$ denotes a random vector which is independently generated from $\mathcal{C} \mathcal{N}(\bm{0}, \bm{I}_{N+1})$. Finally, a candidate  solution to \eqref{attackUMPtest_alternate_vectorform} is given by:
\begin{equation}\label{approximatephaseanglevectorforattack}
{\bm{\Omega}}=\arg \left(\left[\frac{\tilde{\bm{s}}}{\tilde{\bm{s}}(N+1)}\right]_{(1: N)}\right)
\end{equation}
 where $\tilde{\bm{s}}(N+1)$ denotes the $(N+1)$-st element of $\tilde{\bm{s}}$, $[\bm{s}]_{1:N}$ represents the first $N$ elements of the vector $\bm{s}$, and the phases  of all components of the vector $\bm{s}$ are represented by the vector  $\arg(\bm{s})$. Hence, by using \eqref{approximatephaseanglevectorforattack}, a candidate value of $\bm{\Omega}$ is obtained. However, if this candidate $\bm{\Omega}$ does not satisfy all constraints in \eqref{attackUMPtest_alternate_vectorform}, then this solution is discarded.  By generating a  large number of i.i.d.  samples for $\bm{\bar{f}}$ and evaluating only the feasible ones,  with high probability,  we can find a good solution  $\bm{\Omega}$ which nearly minimizes the data rate at the receiver and satisfies the constraints in  \eqref{attackUMPtest_alternate_vectorform};  obviously, this is   a sub-optimal solution\cite{so2007approximating}.
 
 The entire scheme is summarized in Algorithm~\ref{algorithm:For_finding_optimum_attack_phase}.

\begin{algorithm}[H]
	\caption{}
	\label{algorithm:For_finding_optimum_attack_phase}
    \begin{algorithmic}[1]
    \Statex \textbf{Input:} $\bm{g}$,  $\bm{h}$, the parameter $\nu$, a large integer $E$.
    \Statex \textbf{Output:} $\bm{\Omega}$.
    \State Solve \eqref{transformation_to_SDP_form_final} and derive the value of $\bm{S}$.
    \For{$e = 1,2,3, \ldots, E$} 
        \State Sample $\bm{\bar{f}}(e)$ from $\mathcal{C} \mathcal{N}(0, \bm{I}_{N+1})$.
        \State Compute $\bm{\Omega}(e)$  using \eqref{approximatephaseanglevectorforattack}. 
        \State If $\bm{\Omega}(e)$ is a feasible solution of \eqref{attackUMPtest_alternate_vectorform}, evaluate the objective function of \eqref{attackUMPtest_alternate_vectorform} at $\bm{\Omega}(e)$.
    \EndFor
    \State Find   best feasible  solution from $\{\bm{\Omega}(e) \}_{1 \leq e \leq E}$ for  \eqref{attackUMPtest_alternate_vectorform}.
	\end{algorithmic}
\end{algorithm}
\begin{remark}
    We note that the energy detector is UMP, which is the best possible detector for maximizing detection probability under false alarm constraint. Hence, the receiver does not have any incentive to use another detector, irrespective of the phase shift vector $\bm{\Omega}$ used by the attacker. We seek to design the best attack against this UMP detector (see \eqref{attackUMPtest}), where the goal is to minimize the data rate subject to a constraint on the attack detection probability under the UMP detector. One could view \eqref{attackUMPtest} as a conjugate to the problem of finding an attack that minimizes the attack detection probability subject to a constraint on the maximum data rate that the communication is allowed to achieve. Using this new formulation, one can model the interaction between the attacker and the defender  as a constrained zero sum game, where the attacker seeks to minimize the detection probability while restricting the communication data rate, and the detector seeks to maximize the detection probability under a constraint on the false alarm probability. Obviously, the best strategy of the attacker is to use an attack scheme as prescribed by the optimal solution of this new  optimization problem, and the attacker does not have any incentive to use any other $\bm{\Omega}$. Thus,  the optimal solution of this new optimization problem provides a Nash strategy for the attacker, and this strategy along with the energy detector together represent a  Nash equilibrium strategy pair for the zero sum game. In case an alternative to Algorithm~\ref{algorithm:For_finding_optimum_attack_phase}  returns an optimal solution to the new attack design problem, the system can operate at the Nash equilibrium.
\end{remark}
\section{Attack for a given fading block, Quickest Detection under perfect CSI}\label{Sequential Detection}
In this section, we consider a scenario where the receiver detects a possible CMA as the symbols arrive sequentially over a particular fading block. The received symbol under no attack initially follows distribution $f_0$ under hypothesis $H_0$, and owing to CMA at an unknown time, the distribution changes to $f_1$ under hypothesis $H_1$. In our context, the two hypotheses are as follows:
\begin{align*}
H_0 &:  y \sim  \mathcal{CN}(0,\sigma_w^2+\left|\mathbf{g}^{\mathsmaller H} \mathbf{\Phi}_0 \mathbf{h}\right|^{2})  \\
H_1 &:  y \sim  \mathcal{CN}(0,\sigma_w^2+\left|\mathbf{g}^{\mathsmaller H} \mathbf{\Phi} \mathbf{h}\right|^{2})
\end{align*}
After obtaining each sample, the goal of the receiver is to determine whether an attack has already started, and to raise an alarm in case of an attack. We model it as a quickest change detection problem \cite{poor2009quickest,tartakovsky2014sequential,tartakovsky2021asymptotic}. Within a given fading block, the receiver sequentially receives $\{y_{1}, y_{2}, y_3, \cdots \}$. After receiving the $t$-th sample, the receiver decides whether to wait for another sample to detect CMA, or declare right away that a change in received signal statistics has occurred. 
Let $n_c$ represent the time at which the change occurs and $T_{a}$ represent the time at which the receiver raises an alarm. If $T_{a}\geq n_c$, a successful detection occurs and $T_{a}-n_c$ represents the detection delay, whereas if $T_{a} < n_c$, a false alarm event occurs. Obviously, $y \sim  \mathcal{CN}(0,\sigma_w^2+\left|\mathbf{g}^{\mathsmaller H} \mathbf{\Phi}_0 \mathbf{h}\right|^{2})$ if $t<n_c$, and $y_t \sim  \mathcal{CN}(0,\sigma_w^2+\left|\mathbf{g}^{\mathsmaller H} \mathbf{\Phi} \mathbf{h}\right|^{2})$ if $t \geq n_c$.
We are interested in the following three metrics, which will be used to model the detection performance:
\begin{enumerate}
    \item {\em Average Run Length to False Alarm (ARLFA)}   is defined as  $ARLFA \doteq \mathbb{E}_{f_{0}}(T_a)$, {\em i.e.,}  the expected number of samples before a false
    alarm is raised. This quantity must be large.
    \item {\em Average Run Length to Detection (ARLD)}   is the expected number of samples required  to detect a change, {\em i.e.,} $ARLD \doteq \mathbb{E}_{f_{1}}(T_a |n_c=0)$.  ARLD must be as small as possible.
    \item {\em Worst case Detection Delay (WDD)}  for the change detection algorithm is defined as:
    \begin{align}
      W_{T_{a}} \doteq \sup_{n_c \geq 1} \operatorname{ess} \sup \mathbb{E}_{f_{1}}\{(T_{a}-&n_c)^{+} \mid T_{a} \geq n_c,\\
        &\{y_{1},\cdots, y_{n_c} \}\}\nonumber
    \end{align}
\end{enumerate}

\subsection{Attack Detection}

We are interested in non-Bayesian detection of CMA where the detector seeks to solve the following:
\begin{equation}\label{lordens_problem}
    \min_{T_{a}} \hspace{2mm} W_{T_{a}} \quad \text { s.t. } \mathbb{E}_{f_{0}}(T_a)\geq \frac{1}{a_{GLR}}
\end{equation}
where $a_{\mathrm{GLR}}$ is a design parameter based on the required ARLFA,  and is described in detail later. The above detection problem clearly fits into the realm of quickest detection \cite{poor2009quickest,tartakovsky2014sequential} which can   be solved by the   CUSUM  algorithm. 
However,  $\sigma^2 \in [\sigma_{min}^2,\sigma_{0}^2]$ after attack,  where $\sigma_{min}^2$ represents the minimum received signal variance that is achievable by tuning $\bm{\Omega}$, and it can be computed numerically for a given fading block. Since $\sigma^2 \in [\sigma_{min}^2,\sigma_{0}^2]$ after attack, we use the  Generalized Likelihood Ratio CUSUM (GLR-CUSUM) change detection algorithm for  attack detection  \cite{lorden1971procedures}. The GLR-CUSUM algorithm works on the principle of CUSUM algorithm, but additionally involves replacing the likelihood ratio with generalized likelihood ratio. Moreover, GLR-CUSUM algorithm for quickest change detection is asymptotically optimal for   \eqref{lordens_problem} under composite hypotheses \cite[Section~2]{lorden1973open}. The GLR-CUSUM statistic is:
\begin{align}\label{glr_seq}
    \Upsilon_{g}(t)=& \max _{1 \leq k \leq t}\sup _{\sigma^2 \in [\sigma_{min}^2, \sigma_0^2]} \sum_{i=k+1}^{t} l_{G}(y_i)\nonumber\\
    =&\max _{1 \leq k \leq t} \sup _{\sigma^2 \in [\sigma_{min}^2, \sigma_0^2]} \ln \left\{\prod_{i=k+1}^{t}\frac{f_{1,\sigma^2}(y_i)}{f_{0}(y_i)}\right\} \nonumber\\
    =&\max _{1 \leq k \leq t} \sup _{\sigma^2 \in [\sigma_{min}^2, \sigma_0^2]} \sum_{i=k+1}^{t} \left \{ \ln {\frac{{\sigma_0}^2}{{\sigma}^2}}-(\frac{1}{{\sigma}^2}-\frac{1}{{\sigma_0}^2})||y_i||^2 \right \}
\end{align}
 where $ l_{G}(y_i)=\ln (\frac{f_{1,\sigma^2}(y_i)}{f_{0}(y_i)})$ is  the log-likelihood ratio,  and $f_{1,\sigma^2}(y_i)$ represents the probability density function (P.D.F.) of a $\mathcal{CN}(0,\sigma^2)$ random variable. The stopping time of   GLR-CUSUM is defined as:
\begin{equation}
    T_{\mathrm{GLR}}=\inf \left\{t: t \geq 1, \Upsilon_{g}(t) \geq \epsilon_{\mathrm{GLR}} \right\}
\end{equation}
where $\epsilon_{\mathrm{GLR}}$ represents the threshold for the test. Whenever $\Upsilon_{g}(t)$ is greater than the threshold, an alarm is raised. 
Let $\tilde{y}=\sum_{i=k+1}^{t} ||y_i||^2$. Let the supremum in \eqref{glr_seq}  be achieved by $(\sigma^*)^2$. Simple calculation yields:
\begin{equation}\label{seq_sigma_ml}
(\sigma^*)^2= \begin{cases}{\sigma_{0}}^2, & (t-k) \leq \frac{ \tilde{y}}{{\sigma_{0}}^2}, \\ \frac{ \tilde{y}}{t-k}, & \frac{\tilde{y}}{\sigma_{min}^2} \leq(t-k) \leq \frac{ \tilde{y}}{{\sigma_{0}}^2}, \\ \sigma_{min}^2, & (t-k) \geq \frac{ \tilde{y}}{\sigma_{min}^2} .\end{cases}
\end{equation}
The Complex Normal distribution belongs to the exponential family of distributions. Hence, in accordance with \cite[Equation~10]{lai2008quickest} and \cite[Theorem~1]{lorden1973open},  the relationship between the threshold $\epsilon_{\mathrm{GLR}}$, ARLFA, and the lower-bound on ARLD can be expressed as follows: $\epsilon_{\mathrm{GLR}}=- \ln \{a_{\mathrm{GLR}}/b_{\mathrm{GLR}}\}$ where $a_{\mathrm{GLR}}$ is a design parameter based on the required ARLFA, that is 
\begin{equation}\label{seq_arlfa}
    \mathbb{E}_{f_{0}}(T_{\mathrm{GLR}}) \geq 1/a_{\mathrm{GLR}} 
\end{equation}
and moreover from \cite[Theorem~1, Equation~7]{lorden1973open}
\begin{equation}
    \mathbb{E}_{f_{1}}(T_{\mathrm{GLR}}) \geq \frac{\ln a^{-1}_{\mathrm{GLR}}}{I(\sigma^2)}, 
\end{equation}
where ${I(\sigma^2)}$ represents the Kullback–Leibler divergence  ${I(\sigma^2)}=\mathbb{E}_{f_1}\left\{\ln \left\{\frac{f_{1,\sigma^2}(y_i)}{f_{0}(y_i)}\right\}\right\}$, and $b_{\mathrm{GLR}}=3 \ln (a_{\mathrm{GLR}}^{-1})\left(1+\frac{1}{I({\sigma_{min}^2})} \right)^{2}$.
Thus, the receiver can select an appropriate threshold $\epsilon_{\mathrm{GLR}}$ based on \eqref{seq_arlfa}. It can calculate the statistic $\Upsilon_{g}(t)$ in \eqref{glr_seq} using \eqref{seq_sigma_ml}, and compare $\Upsilon_{g}(t)$ to the threshold $\epsilon_{\mathrm{GLR}}$ at each time. When $\Upsilon_{g}(t)$ crosses the threshold, the receiver raises an alarm.
\begin{remark}
It is important to note that, the specific choice for $\epsilon_{GLR}$ was proposed in \cite{lorden1971procedures,lorden1973open} assuming that GLR-CUSUM can receive infinite number of samples, if required. However, in our problem, at most $K$ samples can be transmitted in a given fading block. Hence, the specific choice of $\epsilon_{\mathrm{GLR}}$ in our paper is made under the assumption that $K$ is very large.
\end{remark}
\subsection{Designing the attack against an sequential detector}\label{Designing the attack against an sequential detector}
Now we design an attack strategy against the sequential detector, assuming that the attacker has knowledge of the threshold ${\epsilon_{\mathrm{GLR}}}$. The attacker seeks to find an optimum $\sigma^2$ which is realizable via RIS phase shift, minimizes the data rate at the receiver, and achieves an ARLD greater than a required threshold. The choice of ARLD greater than a certain threshold is motivated by the fact that if the attacker comes to know of the number of samples $K$  transmitted during a fading block, he can simply set $ARLD \geq K$. Mathematically, we can write the optimization problem for the attacker as:
\begin{equation}\label{attacksequential_1}
\begin{aligned}
\min_{\bm{\Omega}} \quad & \log(1+SNR) \hspace{0.4cm}
\textrm{s.t.} \quad ARLD \geq K \hspace{0.3cm}\\
\end{aligned}
\end{equation}
\begin{theorem}\label{theorem:Attack Design against sequential_finding_a}
The  parameter $a_{\mathrm{GLR}}$ for a given    $\epsilon_{\mathrm{GLR}}$ can be calculated by solving the below  equation:
\begin{equation}
    \frac{\ln {a_{\mathrm{GLR}}}}{a_{\mathrm{GLR}}}=-\frac{e^{\epsilon_{\mathrm{GLR}}}}{\bigg\{3\big(1+\frac{1}{I({\sigma_{min}^2})}\big)^{2}\bigg\}} \nonumber \qed
\end{equation}
\end{theorem}
Theorem~\ref{theorem:Attack Design against sequential_finding_a} helps us in deriving the value of parameter $a_{\mathrm{GLR}}$ from the value of threshold, which later helps us in obtaining the constraints for attack design.


\begin{theorem}\label{theorem:Attack Design against sequential_finding_bound_sigma2}
The range of\hspace{1mm} feasible \hspace{1mm} value of \hspace{1mm}$\sigma^2$ \hspace{1mm}for a \hspace{1mm}corresponding lower bound\hspace{1mm} on\hspace{1mm} ARLD \newline (i.e., $\mathbb{E}_{f_{1}}(T_{\mathrm{GLR}}|n_c=0) \geq \tau_{arld}$), where $\tau_{arld}$ is the desired constraint on ARLD, can be approximately obtained by solving the  equation 
   ${{\sigma}^2}-{{\sigma_{0}}^2}\ln ({{\sigma}^2}) \leq L' $, 
where $L'={{\sigma_{0}}^2}(1-\ln{{\sigma_{0}}^2}-\frac{\ln a}{\tau_{arld}})$. The solution of the above equation reduces to ${{\sigma}^2} \geq Q'$ for a suitable threshold  $Q'$. \qed
\end{theorem}

Theorem \ref{theorem:Attack Design against sequential_finding_bound_sigma2} helps us in finding $\sigma^2$ for a given  $\tau_{arld}$. Also, ARLD clearly  increases as $\sigma^2$ increases (i.e., as $|\sigma_0^2-\sigma^2|$ decreases), since it becomes more difficult to distinguish between the two hypotheses.   Theorem \ref{theorem:Attack Design against sequential_finding_bound_sigma2} can be used by an attacker to calculate the phase shift vector $\bm{\Omega}$. However, any arbitrary $\sigma^2$ may not be realizable by $\bm{\Omega}$.  Hence, given that $SNR=\frac{|\bm{{g}}^{H}{\bm{\Phi}\bm{h}}|^{2}}{\sigma_w^2}$, we consider the following   problem instead:
\begin{equation}\label{attacksequential}
\begin{aligned}
\min_{\bm{\Omega}} \quad & \log(1+SNR) \hspace{0.4cm}
\textrm{s.t.} \quad ARLD \geq \tau_{arld} \hspace{0.3cm}\\
\end{aligned}
\end{equation}
\subsubsection{\bf {Procedure to find the attack phase shift vector}}\label{Procedure to find the attack phase shift vector_sequential}
The constraint  $ARLD \geq \tau_{arld}$ can be simplified using Theorem \ref{theorem:Attack Design against sequential_finding_bound_sigma2} as  $\sigma^2 \geq Q' $ or $|\bm{{g}}^{H}{\bm{\Phi}\bm{h}}|^{2} \geq Q'-\sigma_w^2 \doteq \xi_{glr}$. Since $log(\cdot)$ is an increasing function, we can alternatively write \eqref{attacksequential} as:
\begin{equation}\label{attacksequential_alternate_vectorform}
\begin{aligned}
\min_{\bm{\Omega}} \quad & |\bm{{g}}^{H}\bm{{\Phi}{h}}|^{2}\\
\textrm{s.t.} \quad |\bm{{g}}^{H}\bm{{\Phi}{h}}|^{2} \geq \xi_{glr}; \hspace{0.1cm}0&\leq \hspace{0.1cm} \phi_{k}\leq 2\pi, \hspace{0.1cm}1 \leq k \leq N
\end{aligned}
\end{equation}
which is similar to \eqref{attackUMPtest_alternate_vectorform} and thus can be solved using the procedure described in Section~\ref{Procedure to find the attack phase shift vector}.

\section{ Joint beamforming, fixed sample size detector,  multiple transmit antennas}\label{joint design}
In this section, we consider the joint beamforming by the transmitter and RIS in a multiple-input single-output (MISO) communication system where the transmitter consists of $M$ transmit antennas and the RIS is made up of of $N$ elements under continuous phase shift model. We consider independent Rayleigh fading model where the channel between the transmitter and RIS is denoted by the matrix $\bm{G} \in \mathbb{C}^{N \times M}$  and the channel between the RIS and receiver is denoted by $\bm{h}_{r}^{H}\in \mathbb{C}^{1 \times N}$. The beamforming vector at the transmitter is denoted by $\bm{u}_{bm} \in \mathbb{C}^{M \times 1}$, ${x}\sim \mathcal{CN}(0,1)$ represents the transmitted symbol  drawn from a Gaussian codebook, and ${w}\sim \mathcal{CN}(0,\sigma_{w}^2)$ is AWGN. Hence, the received signal is:
\begin{equation}
    y=(\bm{h}_{r}^{H}\bm{\Phi}\bm{G})\bm{u}_{bm}x+w
\end{equation}
The received SNR can be written as $SNR_{jt}\doteq\frac{|\bm{h}^{H}\bm{\Phi}\bm{G}\bm{u}_{bm}|^2}{\sigma_w^2}$. In order  to maximise the received SNR between the transmitter and receiver, the optimal beamforming vector $\bm{u}_{bm}^{*}$ along with the optimal phase shift matrix $\bm{\Phi}_{0}$ is calculated in the RIS during the channel estimation phase, and this vector is fed back to the transmitter for use during data transmission \cite{wu2019intelligent}.   Since the RIS micro controller is under the control of the attacker,  the attacker has knowledge of the channel states as well as $\bm{u}_{bm}^*$. However, the attacker can manipulate the RIS phase shift matrix in the data transmission phase, so that ${y}
\sim \mathcal{CN}(0,\sigma_w^2+ \left|\bm{h}_{r}^{H}\bm{\Phi}\bm{G}\bm{u}_{bm}\right|^{2})$. We define $\sigma^2_{J} \doteq \sigma_w^2+ \left|\bm{h}_{r}^{H}\bm{\Phi}\bm{G}\bm{u}_{bm}^*\right|^2 $ and $\sigma_{J0}^2 \doteq \sigma_w^2+ \left|\bm{h}_{r}^{H}\bm{\Phi}_{0}\bm{G}\bm{u}_{bm}^{*}\right|^2$ as the received signal variances under attack and under no attack.
\begin{lemma}\label{lemma:maxvariance_jt}
$\sigma_{J}^2 \leq \sigma_{J0}^2$ if $\bm{\Phi} \neq \bm{\Phi}_{0}$ or $\bm{u}_{bm}\neq \bm{u}_{bm}^*$.  Equality holds  if $\bm{\Phi} = \bm{\Phi}_{0}$ and $\bm{u}_{bm} = \bm{u}_{bm}^*$ .\qed
\end{lemma}

\subsection{Joint Attack Detection via Hypothesis Testing}
We consider $K$ symbols transmissions during a fading block. The  hypotheses are defined as: 
\begin{align*}
 H_0:  (\bm{\Phi},\bm{u}_{bm})=(\bm{\Phi}_0,\bm{u}_{bm}^*) \hspace{2mm} \text{and} \hspace{2mm} H_1:  (\bm{\Phi},\bm{u}_{bm})\neq (\bm{\Phi}_0,\bm{u}_{bm}^*)
\end{align*}
or equivalently, \
$H_0: \sigma_{J}=\sigma_{J0} \hspace{2mm} \text{and} \hspace{2mm}  H_1:  \sigma_{J} \leq \sigma_{J0}$\\

This problem is is similar to the one discussed in Section~\ref{Attack Detection UMP}, and hence the optimum likelihood ratio test reduces to  an energy detector (see Lemma~\ref{lemma:likelihoodratio}) and is a UMP test (see  Theorem~\ref{theorem:UMP-Test}).
\subsection{Attack design}
Note that the attacker does not change $\bm{u}_{bm}^*$ which could be obtained in many ways including   maximal ratio transmission (MRT \cite{wu2019intelligent}) where  $\bm{u}_{bm}^*=\frac{(\bm{h}_{r}^{\mathsmaller H}\bm{\Phi}\bm{G})^{\mathsmaller H}}{||\bm{h}_{r}^{\mathsmaller H}\bm{\Phi}\bm{G}||}$. 
  The optimization problem for the attacker  becomes: 
\begin{equation}\label{attackUMPtest_jt}
\begin{aligned}
\min_{\bm{\Omega}} \quad & \log(1+SNR_{jt}) \hspace{0.4cm}
\textrm{s.t.} \quad P_{D}\leq\xi_{jt} \hspace{0.3cm}
\end{aligned}
\end{equation}
Now,   we substitute the value of $\bm{u}_{bm}^*$ in the expression for $SNR_{jt}$ and obtain the simplified expression as $\tilde{SNR_{jt}}=\frac{||\bm{h}_{r}^{\mathsmaller H}\bm{\Phi}\bm{G}||^2}{\sigma_{w}^2}$. Since  $P_{D}\leq\xi_{jt}$ can be simplified using Theorem \ref{theorem:Attack Design against UMP} as $\sigma_{J}^2 \geq \frac{{R_{2K,\rho}}{\sigma_{J0}^2}}{R_{2K,\xi}} $ or $||\bm{h}_{r}^{\mathsmaller H}\bm{\Phi}\bm{G}||^{2} \geq (\frac{{R_{2K,\rho}}{\sigma_{J0}^2}}{{R_{2K,\xi}}}-\sigma_w^2 )\doteq \nu_{jt}$, and since $\log(\cdot)$ is an increasing function, we can alternatively write \eqref{attackUMPtest_jt} as follows:
\begin{equation}\label{attackUMPtest_alternate_vectorform_jt}
\begin{aligned}
\min_{\bm{\Omega}} \quad & ||\bm{h}_{r}^{\mathsmaller H}\bm{\Phi}\bm{G}||^{2}\\
\textrm{s.t.} \quad ||\bm{h}_{r}^{\mathsmaller H}\bm{\Phi}\bm{G}||^{2} \geq \nu_{jt}; \hspace{0.1cm}0&\leq \hspace{0.1cm} \phi_{k}\leq 2\pi, \hspace{0.1cm}1 \leq k \leq N
\end{aligned}
\end{equation}
Let $\mathbf{s}=[s_{1}, s_{2} \ldots s_{N}]^{\mathsmaller H}$ where $s_{k}=\mathrm{e}^{j \phi_{k}} $. We denote
$\bm{\Psi}=\operatorname{diag}(\bm{h}_{r}^{\mathsmaller H}) \bm{G}$, we have $||\bm{h}_{r}^{\mathsmaller H}\bm{\Phi}\bm{G}||^{2}=||\bm{s}^{\mathsmaller H}\bm{\Psi}||^2$. Thus, \eqref{attackUMPtest_alternate_vectorform_jt} can be written as:
\begin{equation}\label{transformation_to_SDP_form1_jt}
\begin{aligned}
\min_{\bm{s}} \quad & \bm{s}^{\mathsmaller H}\bm{\Psi}\bm{\Psi}^{\mathsmaller H}\bm{s}\\
\textrm{s.t.} \quad |s_{k}|=1,\hspace{0.1cm}1 \leq &k \leq N;\hspace{0.2cm} \bm{s}^{\mathsmaller H}\bm{\Psi}\bm{\Psi}^{\mathsmaller H}\bm{s}\geq \nu_{jt}
\end{aligned}
\end{equation}
Since \eqref{transformation_to_SDP_form1_jt} is similar to \eqref{transformation_to_SDP_form1}, it can be solved by a similar approach as   in Section~\ref{Procedure to find the attack phase shift vector} and in  Algorithm~\ref{algorithm:For_finding_optimum_attack_phase}.

\section{Attack   over multiple fading blocks under discrete phase shift model}\label{Attack at a large time window}

In this section, we  design the attack  under the discrete phase shift model, though  a similar treatment is possible for the continuous phase shift model as well. We consider single antenna transmitter and a single antenna receiver.   We   assume that the fading in the wireless links is fast and i.i.d. Rayleigh across fading blocks. 
The detector at the receiver collects $\{y_1,y_2,\cdots,y_T\}$   over $T$  fading blocks, using which it has to decide whether an attack has occurred or not. If the detector knows the  channel gains at each time, then it can simply employ an energy detector as in Section~\ref{Attack design against a detector using UMP test}, and the theory for attack detection and design remains unchanged, subject to some scaling operations. However, in this section, we assume that the attacker knows $\{h_k,g_k\}_{1 \leq k \leq N}$ in each fading block, but the receiver does not know it. Hence, the detector seeks to check whether the empirical distribution of SNR over $T$ fading blocks  matches with the SNR distribution under no attack.
This can be done  by performing a goodness-of-fit test, but we consider a simpler detector that only compares   a few empirical moments of received SNR with their desired values; this helps in restricting the number of constraints in the optimization problem of the attacker in Section~\ref{1st attack design}.
After receiving $\{y_1,y_2,\cdots,y_T\}$, the detector computes the empirical moments of SNR, denoted by ${\hat{SNR}}_l$ (for the $l{\text{th}}$ moment), and  declares that an attack has happened if and only if $|{{\hat{SNR}}_l}-{\bar{SNR}_l}| \geq \bar{\zeta}_{l}$, where $\bar{\zeta}_{l}$ represents a threshold and ${\bar{SNR}_l}$ denotes the $l^{\text{th}}$ moment of SNR under no attack. This condition  can be  checked for multiple values of $l$.  In this paper, we focus only on $l=1,2$, though our results can be extended for any $l$. In Section~\ref{Moments of SNR under no attack}, we first derive $\bar{SNR}_l$ in closed form for both continuous and discrete phase shift models, which is later used in Section~\ref{1st attack design} for attack design.
\subsection{Moments of SNR under no attack}\label{Moments of SNR under no attack}
Let us consider Rayleigh channels as discussed in Section~\ref{System Model}, and define the moment generating function (M.G.F) of SNR $\Gamma^{*}$ as   $M_{\Gamma^{*}}(t)=\mathbb{E}(e^{t \Gamma^{*}})$. The $l^{\text{th}}$ moment  of SNR is  $ \bar{SNR}_l=\mathbb{E}\left({\Gamma^{*}}^{l}\right)= \left.\frac{d^{l} M_{\Gamma^{*}}(t)}{d t^{l}}\right|_{t=0} $.
\subsubsection{Continuous phase shift model}
We note from \eqref{receivedsnr2} that the maximum SNR is expressed as $
\Gamma^{*}=\bar{\kappa}|\sum_{k=1}^{N} \alpha_{k} \beta_{k}|^{2}=\bar{\kappa}| Z|^{2}$, 
where $Z=\sum_{k=1}^{N} \alpha_{k} \beta_{k}$.
\begin{theorem}\label{theorem:continuous}
For large number of reflecting elements ($N \rightarrow \infty$), by central limit theorem the coefficient $ Z =\sum_{k=1}^{N} \alpha_{k}  \beta_{k}$ can be approximated by a random variable with distribution  $\mathcal{N}(\mu_Z,\sigma_Z^2)$, where $\mu_{z}=N \sqrt{\epsilon_{h}\epsilon_{g}}\frac{\pi}{4}$,  $\sigma_{z}^2=N\epsilon_{h}\epsilon_{g}(1-\frac{{\pi}^2}{16})$. The SNR can be approximated by a non-central chi-squared distribution with one degree of freedom, also its PDF, M.G.F and moments are expressed as follows:
\footnotesize
\begin{align}
f_{\Gamma^{*}}(\gamma)=&\frac{1}{2\sqrt{2\pi\bar{\kappa}\sigma_{z}^2\gamma}}\bigg({e^{-\frac{{(\sqrt{\frac{\gamma}{\bar{\kappa}}}-\mu_{z}})^2}{2\sigma_{z}^2} }}+{e^{-\frac{{(\sqrt{\frac{\gamma}{\bar{\kappa}}}+\mu_{z}})^2}{2\sigma_{z}^2}}}\bigg)\nonumber\\
M_{\Gamma^{*}}(t)=&e^{\mu_Z^2\bar{\kappa} t /(1-2\bar{\kappa} t\sigma_Z^2)}(1-2\bar{\kappa} t\sigma_Z^2)^{-\frac{1}{2}}\nonumber\\
\mathbb{E}(\Gamma^{*})=&N\bar{\kappa}\epsilon_{h}\epsilon_{g}\bigg(1+\frac{\pi^2}{16}(N-1)\bigg)\nonumber\\
\mathbb{E}({\Gamma^{*}}^2)=&N^2\bar{\kappa}^2\epsilon_{h}^2\epsilon_{g}^2\bigg(1+\frac{\pi^4}{256}(N-1)^2 + \frac{\pi^2}{8}(N-1)+\nonumber\\
    &\frac{16-\pi^2}{8}+\frac{\pi^2(2N-1)(16-\pi^2)}{128}\bigg)\nonumber
\end{align}
\normalsize
\end{theorem}

\subsubsection{Discrete phase shift model}
 By  \eqref{receivedsnr2}, the SNR received under discrete phase shift model is:
\begin{equation}\label{discretereceivedSNR}
    \Gamma^{*}=\bar{\kappa}\left|\sum_{k=1}^{N} \alpha_{k} \beta_{k}e^{j \delta_{k}}\right|^{2}=\bar{\kappa}{N}^2|V|^{2}
\end{equation}
where $V \doteq \frac{1}{N}\sum_{k=1}^{N}{\alpha_k}{\beta_k}e^{j\delta_{k}}$. Let us define characteristic function of $\delta_k$ as  $\varphi_{\delta_k}(\omega) \doteq \mathbb{E}[e^{j\omega\delta_k}]$,  $\omega \in \mathbb{R}$. 
\begin{theorem}\label{theorem:discrete}
For large number of reflecting elements ($N \rightarrow \infty$), by central limit theorem $V=V_R+ j V_I$ can be approximated to a complex Gaussian distribution. The real and imaginary parts $V_R$ and $V_I$ are independent with $V_R \sim \mathcal{N}(\mu,\sigma_{V_R}^2)$ and $V_I \sim \mathcal{N}(0,\sigma_{V_I}^2)$, where $\mu=\frac{\pi}{4}\sqrt{\epsilon_{h}\epsilon_{g}}\varphi_{\delta_k}(1)$, $\sigma_{V_R}^2=\frac{{\epsilon_h}{\epsilon_g}}{2N}(1+\varphi_{\delta_k}(2)-\frac{\pi^2}{8}{\varphi_{\delta_k}(1)}^2)$ and $\sigma_{V_I}^2=\frac{{\epsilon_h}{\epsilon_g}}{2N}(1-{\varphi_{\delta_k}(1)})$. Also, defining $\theta_{|V|^{2}} \doteq 4\sigma_{V_R}^2$ and $k_{|V|^{2}} \doteq \tfrac{\mu^2}{4\sigma_{V_R}^2} $, the SNR can be approximated by a Gamma distribution with its PDF, M.G.F and  moments expressed as follows:
\begingroup\makeatletter\def\f@size{10}\check@mathfonts
\def\maketag@@@#1{\hbox{\m@th\large\normalfont#1}}%
\footnotesize
\begin{align}
f_{\Gamma^{*}}(\gamma)=&{\frac {\gamma^{{k_{V^{2}}}-1}e^{-\frac{\gamma}{\bar{\kappa}{N}^2\theta_{V^{2}}}}}{{(\bar{\kappa}{N}^2\theta_{V^{2}})} ^{k_{V^{2}}}\Gamma (k_{V^{2}})}}\nonumber\\
    M_{\Gamma^{*}}(t)=&\big(1-t\varrho \theta_{|V|^{2}}\big)^{-k_{|V|^{2}}}\nonumber\\
    \mathbb{E}(\Gamma^{*})=&\frac{N^2\bar{\kappa} \pi^2}{16}\epsilon_h\epsilon_g{\varphi_{\delta_k}(1)}^2\nonumber\\
    \mathbb{E}({\Gamma^{*}}^2)=&\epsilon_h^2\epsilon_g^2\bigg(\frac{N^3 \pi^2}{4}\bigg(1+{\varphi_{\delta_k}(2)}-\frac{\pi^2 {\varphi_{\delta_k}(1)}^2}{8}\bigg)+  
    \frac{N^4\pi^4{\varphi_{\delta_k}(1)}^4}{256}\bigg)\nonumber
\end{align}
\normalsize
\endgroup

\end{theorem}
Using Theorem \ref{theorem:discrete} we can find out any moment of SNR under the discrete phase shift model.

\subsection{Attack design: optimization problem formulation under  discrete phase shift model}\label{1st attack design}
Let  $s$ be a typical state   of the attacker, which consists of the instantaneous  channel gains $\{h_k,g_k\}_{1 \leq k \leq N}$ available to the attacker.  However, we discretize the state space to obtain a finite number of states; this is reasonable as long as the granularity of discretization is small enough to approximate Raylegh fading gain's PDF accurately.  However, the formulation in this subsection can also be extended to continuous values channel gains, where certain summations in the next optimization problem become integrations. 

A generic action $a$ of the attacker is a vector of phase shifts  $[\hat{\phi_{1}},\ldots,\hat{\phi_{N}}] \in \mathcal{D}^N$; the attacker takes an action in each fading block. 
The signal to noise ratio under a state-action pair is denoted by $SNR(s,a)$, and this can be computed by the attacker using \eqref{receivedsnr2}. Moreover,  ${\bar{SNR}_l}$ can be calculated by an attacker using either Theorem \ref{theorem:discrete} for large $N$ or by brute-force computation for small $N$. The  attacker seeks to minimise the ergodic data rate at the receiver while roughly preserving the moments of SNR. In order to allow some tolerance margin, the attacker uses a threshold  $\zeta_{l} \doteq \kappa_{l} \bar{\zeta}_{l}$ instead of $\bar{\zeta}_l$ for attack design, where $\kappa_{l}$ is a proportionality constant quantifying the tolerance level. 

Let $\pi(s)$ denote the probability of occurrence of   channel state $s$. We consider probabilistic action selection by the RIS at each time; let $p(a|s)$ denote the probability of choosing action $a$ under  state $s$ on part of the attacker. 
The attacker seeks to minimize the ergodic data rate at the receiver,  while preserving the moments of the received SNR. The attacker's problem can be cast as follows:
\begin{equation}\label{attackoptimizationproblem}
\begin{aligned}
\min_{p(a|s)_{\forall a,s}}\quad &\sum_{s} \pi(s)\sum_{a}p(a|s)\log(1+SNR(s,a))\\
\textrm{s.t.} \hspace{+4mm}\quad  &p(a|s)\geq 0 \hspace{0.2cm}\forall a,s;\hspace{0.1cm} \sum_{a}p(a|s)=1 \hspace{0.2cm}\forall s\\
 |\sum_{s} \pi(s)&\sum_{a} p(a|s)(SNR(s,a))-\bar{SNR}_l|\leq \zeta_{l}
\end{aligned}
\end{equation}
Let us consider $m$ actions and $n$ states. We denote
$\mathbf{p} \doteq [\mathbf{p_1},\ldots,\mathbf{p_k},\ldots,\mathbf{p_n}]^{\mathsmaller T}$, where $\mathbf{p_k} \doteq [\pi(s_{k}) \bm{1}_{m}]^{\mathsmaller T}$, and $\bm{1}_{m}$ is an all-$1$ column vector of dimension $m \times 1$. Let  $\mathbf{x} \doteq [\mathbf{x_1},\ldots,\mathbf{x_k},\ldots,\mathbf{x_n}]^{\mathsmaller T}$, where \hspace{2mm} $\mathbf{x_k}  \doteq [p(a_{1}|s_{k}),\hspace{2mm}\ldots\hspace{2mm}, p(a_{m}|s_{k})]$.\hspace{2mm} Also,\hspace{2mm} $\mathbf{C}\hspace{2mm}\doteq\hspace{2mm}\operatorname{diag}([{\mathbf{t_{1}}},\hspace{2mm}\ldots,\mathbf{{t_{k}}},\ldots,\mathbf{{t_{n}}}])$  where \\ $\mathbf{{t_{k}}}=[SNR(s_k,a_1),\hspace{2mm}\ldots, \hspace{2mm} SNR(s_k,a_m)]$,\hspace{2mm} $\mathbf{K}\doteq\operatorname{diag}([\mathbf{{u_{1}}},\hspace{2mm}\ldots,\hspace{2mm}\mathbf{{u_{k}}},\hspace{2mm}\ldots,\hspace{2mm}\mathbf{{u_{n}}}])$,\hspace{2mm} where \\ $\mathbf{{u_{k}}}=[\log(1+SNR(s_{k},a_{1})),\ldots,\log(1+SNR(s_{k},a_{m}))]$, and $\mathbf{R} \doteq \mathbf{C}^2$. The optimization problem   \eqref{attackoptimizationproblem} can be rewritten as a linear optimization problem as follows:
\begin{equation}\label{linearattackoptimizationproblem}
\begin{aligned}
\min_{\mathbf{x}} \quad & \mathbf{p}^{\mathsmaller T}\mathbf{K} \mathbf{x}\\
\textrm{s.t.} \quad \mathbf{x} \geq 0,\hspace{0.1cm}  &\mathbf{A}\mathbf{x}={\bm{1}_{n \times 1}},  \hspace{0.3cm} \\
  \mathbf{p}^{\mathsmaller T}\mathbf{C} \mathbf{x} \leq \bar{SNR}_{1} + \zeta_{1} &,\hspace{0.1cm}
  \mathbf{p}^{\mathsmaller T}\mathbf{C} \mathbf{x} \geq \bar{SNR}_{1} - \zeta_{1}\\
  \mathbf{p}^{\mathsmaller T}\mathbf{R} \mathbf{x} \leq \bar{SNR}_{2} + \zeta_{2} &,\hspace{0.1cm}
  \mathbf{p}^{\mathsmaller T}\mathbf{R} \mathbf{x} \geq \bar{SNR}_{2} - \zeta_{2}\\
\end{aligned}
\end{equation}
The matrix $\mathbf{A}$ is a logical matrix used to represent the constraint $\sum_{a}{p(a|s)}=1 \hspace{0.2cm}\forall s$, and its entries are chosen accordingly. This linear program can be solved by using traditional solvers, and its solution can be used by the attacker.

Note that, for continuous phase shift model, similar optimization problems can be formulated, either by changing the summations to integrations or by discretizing the state and action spaces.  

\section{Attack in a given fading block, fixed sample size detector, imperfect CSI}\label{CSI uncertainty}
In this section, we examine channel uncertainty  under the continuous phase shift model. We consider a single antenna transmitter and receiver, and a RIS. Due to the passive nature of the RIS, it is difficult to obtain the vector $\bm{\Psi}=\operatorname{diag}(\bm{g}^{H}) \bm{h}$ and the cascaded scalar channel gain $\bm{s}^{\mathrm{H}}{\bm{\Psi}}$ at the receiver and RIS, where $\bm{s}=[s_{1}, s_{2} \ldots s_{N}]^{\mathsmaller H}$ and $s_{k}=\mathrm{e}^{j \phi_{k}} $ denotes the phase shifts of the RIS element. Hence, we consider the statistical CSI error model   \cite{hong2020robust} to model the channel estimation error. We assume that   $\bm{\Psi}$   is inaccurately estimated, which can be expressed as:
\begin{equation}
    \bm{\Psi}=\hat{\bm{\Psi}}+ \Delta\bm{\Psi}
\end{equation}
where $\hat{\bm{\Psi}}=[\hat{\psi_{1}}, \hat{\psi_{2}} \ldots \hat{\psi_{N}}]$ is the estimated cascaded CSI and $\Delta\bm{\Psi}\sim \mathcal{C} \mathcal{N}(\bm{0}, {\sigma_{e}^2}\bm{I})$ is the unknown cascaded CSI error with known distribution. We can choose  ${\sigma_{e}^2}$ in accordance with the worst case CSI uncertainty scenario. We consider a quasi-static flat fading block under consideration, $P=1$ and continuous phase shift model for the RIS. The transmitted symbols are drawn from a Gaussian codebook, and  hence ${x} \sim \mathcal{CN}(0,1)$, $w \sim \mathcal{CN}(0,\sigma_w^2)$.
The signal $y_{uc}$ at the receiver   can be expressed as follows using \eqref{Composite channel}:
\begin{equation}\label{Received_signal_CSI_uncertainty}
    y_{uc}=\bm{s}^{\mathrm{H}}{\bm{\Psi}}x+w=\bm{s}^{\mathrm{H}}( \hat{\bm{\Psi}}+ \Delta\bm{\Psi}) x+w
\end{equation}
where $\bm{s}=[s_{1}, s_{2} \ldots s_{N}]^{\mathsmaller H}$ denotes the vector consisting of the RIS phase shifts, with  $s_{k}=\mathrm{e}^{j \phi_{k}}$. The receiver and the RIS know the estimate $\hat{\bm{\Psi}}$ but are unsure about $\Delta\bm{\Psi}$, hence the RIS sets its phase shift in accordance with  $\hat{\bm{\Psi}}$, namely $s_{k}=\mathrm{e}^{-j \arg(\hat{\psi_{k}})},\hspace{1mm} \forall \,\, 1\leq k\leq N$.

\begin{theorem}\label{theorem:pdf_under_CSIR_uncertainty}
The probability density function   of the received signal $y_{uc}$ under channel uncertainty is: 
\begin{align*}
f_{y_{uc}}(y_{uc})&=\frac{1}{\pi\sigma_{s}^2} \exp (\frac{\sigma^{2}}{\sigma_{s}^2}) \sum_{i=0}^{\infty}(\frac{1}{\sigma_{s}^{2} })^{i} G(-i, \frac{\sigma^{2}}{\sigma_{s}^{2}}) \frac{(-{y_{uc}}^{2})^{i}}{i!}
\end{align*}
where  $\sigma^{2} = \sigma_{0}^2 \doteq \sigma_{w}^{2} + |\sum_{i=0}^{N}|\hat{\psi_{i}}||^2$ under no CMA, and  $\sigma^{2}=\sigma_{w}^{2} + |\sum_{i=0}^{N}\hat{\psi_{i}}\mathrm{e}^{j \phi_{i}}|^2$ under CMA. Here $\sigma_{s}^{2}=N{\sigma_{e}}^2$, and $G(\cdot, \cdot)$  denotes the incomplete Gamma function \cite{harris2008incomplete}. Moreover, the received signal energy   ${r}=|y_{uc}|^2$ has  its cumulative distribution function (C.D.F) expressed as follows:
\begin{align*}
F_{{r,\sigma}}(r)&=1-[\frac{\sigma^{2}}{\sigma_{s}^{2}}\mathrm{e}^{-(\frac{\sigma^{2}}{\sigma_{s}^{2}})} \int_{1}^{\infty}(\mathrm{e}^{(-\frac{\sigma^{2}t}{\sigma_{s}^{2}}-\frac{r}{\sigma^{2}t})})\,dt   ] \qed
\end{align*}
\end{theorem}
Since the P.D.F. of received signal is   intractable for likelihood ratio test,   we consider a non-parametric detection technique namely Kolmogorov–Smirnov test (KS test) on the energy of received signal.


\subsection{Attack Detection using KS-Test}
The KS test is a non-parametric goodness of fit test \cite[Chapter~6, Page~430]{conover1999practical}. We denote    the empirical C.D.F of energy of the received samples by $\hat{F_{1}}$, and  C.D.F under no attack (obtained using Theorem~\ref{theorem:pdf_under_CSIR_uncertainty}) by $F_{r,\sigma_0}$. The hypotheses are represented as follows:
\begin{align*}
H_0: \hat{F_{1}}=F_{r,\sigma_0} \hspace{8mm} \text{and} \hspace{8mm}
H_1: \hat{F_{1}} \neq F_{r,\sigma_0}
\end{align*}

For $K$ received samples with sample energies $\{r_1, r_2, \cdots, r_K \}$, the test statistic   $\bar{D}$ is written as:
\begin{align}
\bar{D} \doteq \max _{1 \leq n \leq K}\left|\hat{{F}_{1}}\left(r_{n}\right)-F_{r,\sigma_{0}}\left(r_{n}\right)\right| 
\end{align}
The test statistic is compared with a critical value denoted by  $\bar{\iota}$ obtained from the K-S table \cite[Table~A13, Page~547]{conover1999practical} corresponding to a required significance level $\hat{\alpha}$. The test can be expressed as follows:
\begin{equation}
    \bar{D} \hspace{4mm} 
    {\underset{H_{1}}{\overset{H_{0}}{\lesseqqgtr}}} \hspace{4mm} \bar{\iota}
\end{equation}

  Let $\hat{D}_n \doteq\hat{{F}_{1}}(r_{n})-F_{r,\sigma_0}(r_{n})$.  The KS test raises an alarm if $|\hat{D}_n| \geq \bar{\iota}$, for at least one $n, \hspace{1mm} 1 \leq n \leq K$. This is a per-sample test. 
Since $r_n, 1 \leq n \leq K$, is i.i.d under no attack,  from the detector's perspective, the posterior probability  $\mathbb{P}(|\hat{D}_n| \geq \bar{\iota}  |\  r_n, H_0) $ needs to be less than a  small positive number $\bar{\varepsilon}_{ks}$ in order to achieve a low false alarm probability. 

\begin{theorem}\label{theorem:KS_Test_detector_energy_range}
For a given threshold $\bar{\iota}>0$, the constraint   $\mathbb{P}(|\hat{D}_n| \geq \bar{\iota}  |\  r_n, H_0) \leq \bar{\varepsilon}_{ks}$ requires either $r_n \geq r_{u}$ or $r_n \leq r_{l}$, where $r_{u}$ and $r_{l}$ denote the upper and lower thresholds on the energy of the received samples, respectively. Also, the value of the two thresholds  can be approximately calculated by  the expressions $r_{u}=F^{-1}_{r,\sigma_0}\bigg(\frac{1+\sqrt{1-4(K-1)\bar{\varepsilon}_{ks} {{\bar{\iota}^2}}}}{2}\bigg)$ and $r_{l}=F^{-1}_{r,\sigma_0}\bigg(\frac{1-\sqrt{1-4(K-1)\bar{\varepsilon}_{ks} {{\bar{\iota}^2}}}}{2}\bigg)$, provided that $\bar{\varepsilon}_{ks}$ is small enough to ensure that $r_l$ and $r_u$ are real numbers. \qed
\end{theorem}
We note that $r_l$ and $r_u$ can be computed by inverting the CDF $F_{r,\sigma_0}(\cdot)$ defined in  Theorem~\ref{theorem:pdf_under_CSIR_uncertainty}.
Theorem~\ref{theorem:KS_Test_detector_energy_range} provides us a possible range of $r_n$, which helps us in attack design.

It is interesting to note that Theorem~\ref{theorem:KS_Test_detector_energy_range} yields a {\em per-sample double threshold rule} to find the range of $r_n$ that ensures  $\mathbb{P}(|\hat{D}_n| \geq \bar{\iota}  |\  r_n, H_0) \leq \bar{\varepsilon}_{ks}$. It is obvious that the detector can simply raise an alarm if $r_n \leq r_l$ or $r_n \geq r_u$, and this will result in a  small false alarm probability. However, this alternative detection rule seems to be counter-intuitive since one would expect the opposite to happen in case of the simple energy detector. This phenomenon can be explained if we take a closer look at the KS test. The KS-test and the subsequent per sample variation of it always raise an alarm when the difference between two CDFs crosses a threshold, and this is more probable when the sample is neither too large nor too small. On the other hand, a large value of $r_n$ will usually result in a smaller KS statistic and hence will be treated as a typical sample by the KS test.

\subsection{Attack Design}\label{AttackDesignImperfect}
Now we  design an attack strategy against the per sample double threshold test, assuming that the attacker has knowledge of $\bar{\iota}$.   Since the attacker seeks to reduce the data rate to the user, it does not make any sense to the attacker to adjust its induced phase shift and result in a large value in received sample energy $r_n$. Hence, the attacker will try to ensure  $\mathbb{P}(r_n > r_l|H_1) \leq \nu_{ks}$ for a  small $\nu_{ks}>0$, which holds if $\mathbb{E}(r_n|H_1) \leq \nu_{ks} r_l$ (by Markov inequality). Since $\mathbb{E}(r_n|H_1)=|\bm{s}^{H}\bm{\hat{\Psi}}|^2+N \sigma_e^2+\sigma_w^2$ and SNR at the receiver under attack is $\frac{|\bm{s}^{H}\bm{\hat{\Psi}}|^2+N \sigma_e^2}{\sigma_w^2}$, we can write the optimization problem for the attacker as follows:
\begin{equation}\label{Unknown_CSI_opti}
\begin{aligned}
   \min_{\bm{s}} \quad & \bm{s}^{H}\bm{\hat{\Psi}}\bm{\hat{\Psi}}^{H}\bm{s}\\
\textrm{s.t.}\hspace{4mm} |s_{k}|=1,\hspace{1mm}1 \leq &k \leq N;
\hspace{2mm}\bm{s}^{H}\bm{\hat{\Psi}}\bm{\hat{\Psi}}^{H}\bm{s} \leq \nu_{ks} r_l-N \sigma_e^2 -\sigma_w^2
\end{aligned}
\end{equation}
However, if the attacker is also cautious about  a possible energy detector, one additional constraint $\mathbb{E}(r_n|H_1) \geq \tilde{\nu}_{ks}'$ or equivalently  $\bm{s}^{H}\bm{\hat{\Psi}}\bm{\hat{\Psi}}^{H}\bm{s} \geq \tilde{\nu}_{ks}'-N \sigma_e^2 -\sigma_w^2 \doteq \tilde{\nu}_{ks}$ can be incorporated in \eqref{Unknown_CSI_opti}. Both versions of this   problem can be solved by using similar procedures as in   Section~\ref{Procedure to find the attack phase shift vector}.

\subsection{Quickest detection of CMA over a given fading block under imperfect CSI} 
Under imperfect CSI, the distribution of the received symbols before and after attack can be characterised by the P.D.F. as obtained in closed form in Theorem~\ref{theorem:pdf_under_CSIR_uncertainty}. Hence, we can theoretically use the GLR-CUSUM algorithm  as mentioned in Section~\ref{Sequential Detection} for quickest detection of CMA over a given fading block under imperfect CSI. However, due to the complex expression in Theorem~\ref{theorem:pdf_under_CSIR_uncertainty}, GLR-CUSUM may not yield any simple solution, and    non-parametric sequential change point detection algorithms can be used instead  \cite{shin2022detectors}. Analysing such algorithms is beyond the scope of this paper, hence we leave it for future research.

\subsection{Discussion for non-stationary channel}
While the channel statistics is usually stationary for static systems, it will be non-stationary when the receiver is highly mobile. Channel gain can be known at the receiver so long as the transmitter sends pilot symbols at the beginning of every fading block. Of course, this can result in channel estimation error, which has already  been addressed in this section. 
If the channel statistics is non stationary but known at every fading block, and an erroneous estimate of the channel gain is available in each fading block, then the analyses provided earlier in this section will still work. However, the linear program formulation for attack design over multiple fading blocks as in Section~\ref{Attack at a large time window} will not be valid because the channel does not have a stationary distribution. In this case, if the channel statistics is non stationary but slowly varying, such that the variation is channel statistics can be bounded above by a variation budget, then a conservative linear program can be formulated to design the attack against the worst possible variation of channel statistics over a finite number of fading blocks. Formulating this problem and solving it is beyond the scope of this paper.
If channel gains are unknown, and channel statistics is also non stationary across fading blocks and unknown, then one can focus on  quickest detection over a single fading block (Section~\ref{Sequential Detection}) and detection over multiple fading blocks (Section~\ref{Attack at a large time window}). Under such setting, one has to resort to non stationary detection techniques, which can be tried in future research on this topic.

\section{Numerical Results}\label{NumericalResults}
\noindent We consider Rayleigh fading model, i.e., $h_{k}, g_{k} \sim \mathcal{CN}(0,1)$. Also, $P=30$ dBm,  $\sigma_{w}^2=-10$ dBm. We compare our attack algorithms with a baseline that chooses   RIS phase shifts uniformly at random. 
\begin{figure}[!tbp]
  \centering
  \begin{minipage}[b]{0.4\textwidth}
    \includegraphics[width=\textwidth]{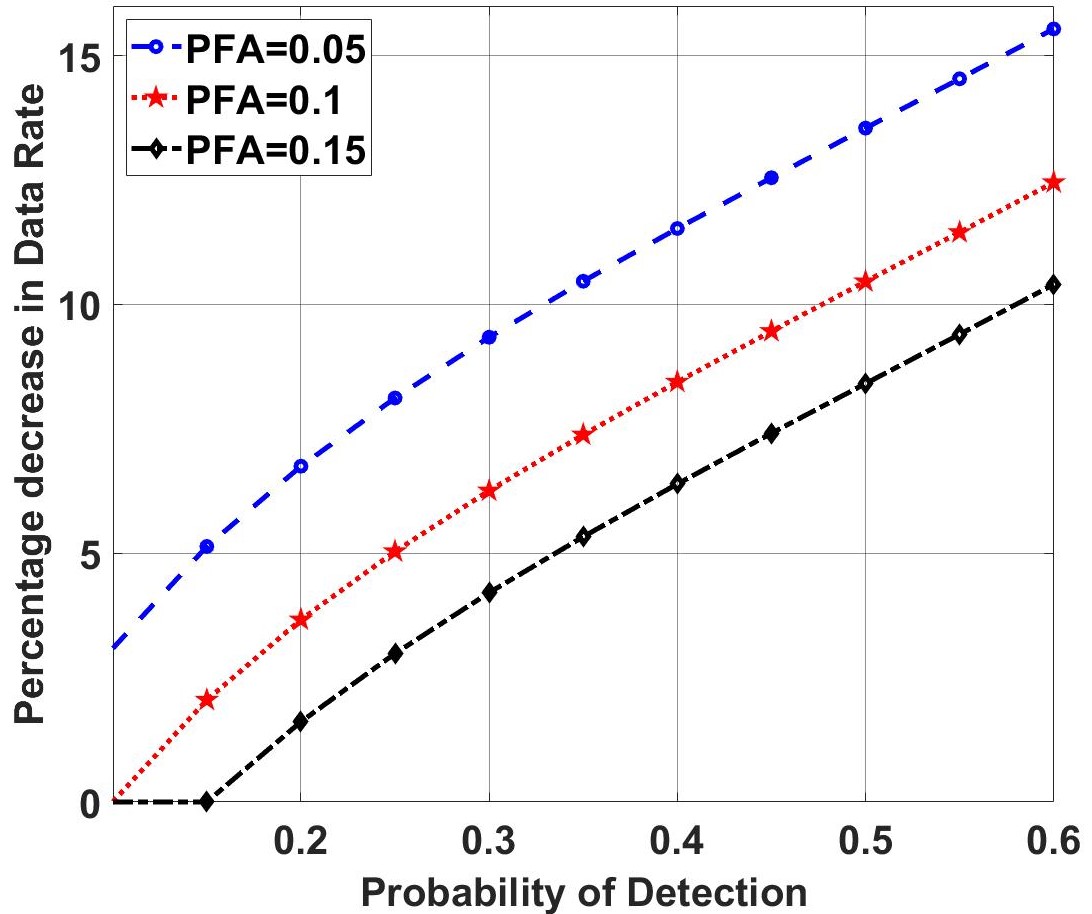}
    \caption{Plot for percentage decrease in data rate for different false alarm probabilities in case of UMP Test.}\label{fig:Decrease in data rate vs prob of detection in ump testing}
  \end{minipage}
  \hfill
  \begin{minipage}[b]{0.4\textwidth}
    \includegraphics[width=\textwidth]{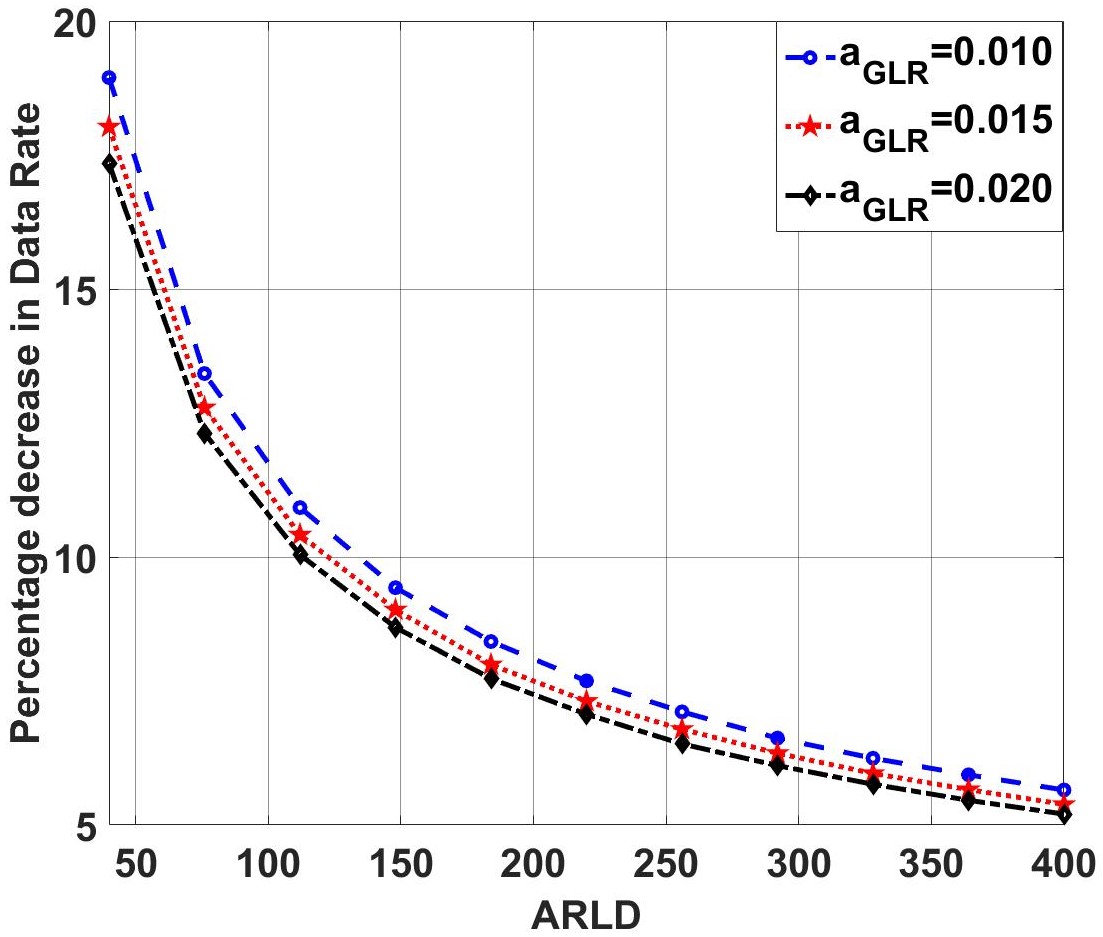}
    \caption{Percentage decrease in data rate versus ARLD achieved by an attacker against    GLR-CUSUM.}\label{fig:Percentage Decrease with ARLD}
  \end{minipage}
\end{figure}

\begin{table}[ht!]
\Huge

\centering\caption{{Comparison with baseline scheme in the scenario described in Section~\ref{Attack design against a detector using UMP test} }}
 \label{tab:Table_1_baseline_UMP}
\resizebox{8.7cm}{!}{\begin{tabular}{|c|c|c|c|}
 \hline
 \begin{tabular}{@{}c@{}} Probability of \\false alarm\\ \end{tabular} &
 \begin{tabular}{@{}c@{}}Expected \\probability \\ of detection \end{tabular}&\begin{tabular}{@{}c@{}}Percentage decrease \\in data rate \\for random phase \\shift attack model\\ \end{tabular}&\begin{tabular}{@{}c@{}}Percentage decrease \\in data rate \\for our algorithm \end{tabular}  \\
 \hline
 \begin{tabular}{@{}c@{}}PFA=0.05\\ \end{tabular} & 0.53& 9.1\%&14.1\%\\
 \hline
 \begin{tabular}{@{}c@{}}PFA=0.10\\ \end{tabular} & 0.57&7.56\%&12.6\%\\
 \hline
 \begin{tabular}{@{}c@{}}PFA=0.15\\ \end{tabular} & 0.60& 6.40\%&11.2\%\\
 \hline
 \end{tabular}}
 \end{table}

\subsection{UMP test over a given fading block, fixed sample size detector}\label{results_UMP}
We consider $N=64$  and  $K=50$, and evaluate the performance of the proposed attack scheme averaged over $10000$ independently generated channel realizations. For various false alarm rates, we evaluate the threshold ($\eta'$) and use it to find  $\sigma^2$  for a given probability of detection.  Further, using Algorithm \ref{algorithm:For_finding_optimum_attack_phase}, we find out the optimum phase shift matrix ($\bm{\Omega}$) and employ it to attack the generated instances.  Fig \ref{fig:Decrease in data rate vs prob of detection in ump testing} shows that,  as the probability of detection is increased, the attacker can decrease the data rate by a considerably large value.  Additionally,   as the probability of false alarm increases for a fixed detection probability, the percentage decrease in data rate    reduces. This is due to the fact that, as the false alarm rate  increases, the detector becomes more cautious about the attack, and   leaves limited opportunity  for the attacker to reduce the data rate. 
We have observed from multiple simulations that our attack algorithms performs much better under low SNR. The intuition is that under low SNR, data rate $\log(1+SNR) \approx SNR$ and the receiver receives poor signal strength; hence, slight change in SNR caused by the attack results in high percentage decrease in the data rate. On the other hand, $\log(1+SNR) \approx \log (SNR)$ in the high SNR regime, and the $\log$ function makes the data rate less sensitive to the attack.

The baseline technique (random phase shift attack) is compared with our algorithm in Table~\ref{tab:Table_1_baseline_UMP}. For a given value of false alarm probability and given channel realization, we first compute the detection threshold $\eta'$ using Theorem~\ref{theorem:UMP-Test}. Next,  we simulate the detection probability and mean data rate under random phase shift attack for an energy detector using this threshold $\eta'$.  By averaging over many  channel realizations, we calculate the  expected attack detection probability $\tilde{P}_D$ and  the expected percentage decrease in data rate under random phase shift attack. Next, for each channel realization, we set  $\xi=\tilde{P}_D$  in \eqref{attackUMPtest} to find the optimal $\bm{\Omega}$; this is done by computing  $\sigma^2$ from Theorem~\ref{theorem:Attack Design against UMP}, using it in \eqref{attackUMPtest_alternate_vectorform}, and running Algorithm~\ref{algorithm:For_finding_optimum_attack_phase}. We average the results over multiple independent channel realizations to find the expected percentage decrease of data rate under our attack scheme. 
Table~\ref{tab:Table_1_baseline_UMP} shows that our attack scheme results in significantly larger percentage decrease in data rate compared to random phase shift attack, for the same false alarm probability and expected attack detection probability. 
Also, we notice that the attack detection probability under random phase shift attack is very large and it can not be reduced since $\eta'$ is determined by the false alarm probability which is chosen by the receiver. Our algorithm provides optimal attack in this case.

\begin{table}[ht!]
\Huge
\centering
\centering\caption{Comparison with random phase shift attack scheme for the scenario described in Section~\ref{Sequential Detection} }
 \label{tab:Table_1_baseline_sequential}
\resizebox{8.0cm}{!}{\begin{tabular}{|c|c|c|c|}
\hline
 \begin{tabular}{@{}c@{}} Values of \\parameter $a_{\mathrm{GLR}}$\\ \end{tabular} &
 \begin{tabular}{@{}c@{}}Expected \\average run length\\to detection \end{tabular}&\begin{tabular}{@{}c@{}}Percentage decrease \\in data rate \\for random phase \\shift attack model\\ \end{tabular}&\begin{tabular}{@{}c@{}}Percentage decrease \\in data rate \\for our algorithm \end{tabular}  \\
 \hline
 \begin{tabular}{@{}c@{}}$a_{\mathrm{GLR}}=0.01$\\ \end{tabular} & 137&6.58\%&10.1\%\\
 \hline
 \begin{tabular}{@{}c@{}}$a_{\mathrm{GLR}}=0.015$\\ \end{tabular} & 185&6.3\%&7.8\%\\
 \hline
 \begin{tabular}{@{}c@{}}$a_{\mathrm{GLR}}=0.020$\\ \end{tabular} & 188&6.1\%&7.3\%\\
 \hline
 \end{tabular}}
\end{table}

\subsection{Sequential detection over a given fading block}
We consider an RIS with $N=64$ elements to simulate the attack described in Section~\ref{Sequential Detection}. In Fig \ref{fig:Percentage Decrease with ARLD}, we plot the performance of the proposed attack scheme, averaged over $10000$ independently generated channel realizations. For various values of the parameter $a_{\mathrm{GLR}}$, we evaluate the value of the threshold $\epsilon_{\mathrm{GLR}}$. Next,  for a given value of ARLD, we evaluate the value of $\sigma^2$ using Theorem ~\ref{theorem:Attack Design against sequential_finding_a} and Theorem~\ref{theorem:Attack Design against sequential_finding_bound_sigma2}. Then, using Algorithm \ref{algorithm:For_finding_optimum_attack_phase}, we compute the optimal phase shift matrix $\bm{\Omega}$ and employ it to attack the generated instances.  Fig \ref{fig:Percentage Decrease with ARLD} shows    the percentage decrease in data rate decreases with ARLD under this attack; this conforms with the intuition that,  as the value of ARLD becomes larger, the attacker needs to maintain a $\sigma^2$   close to $\sigma_{0}^2$ to avoid getting detected. Also, as  $a_{\mathrm{GLR}}$ increases, ARLFA decreases, and hence $\epsilon_{\mathrm{GLR}}$ decreases. This results in lower ARLD for the same percentage decrease in data rate.
Furthermore, as  in Section~\ref{results_UMP}, here also we observe that our algorithm performs much better in low SNR regime. 
Table~\ref{tab:Table_1_baseline_sequential} compares  our attack design algorithm  with the baseline scheme (random phase shift attack).  For a given  ARLFA, we first compute the  parameter $a_{\mathrm{GLR}}$ using \eqref{seq_arlfa}. Next, for a given  $a_{\mathrm{GLR}}$ and given channel realizations, we   compute the detection threshold $\epsilon_{\mathrm{GLR}}$ by using Theorem~\ref{theorem:Attack Design against sequential_finding_a}. Subsequently, we simulate the ARLD and mean data rate under random phase shift attack for the GLR-CUSUM detector using the threshold $\epsilon_{\mathrm{GLR}}$. By averaging over various channel realizations we calculate the expected ARLD ($\tilde{ARLD}$) and expected percentage decrease in data rate under random phase shift attack. Next, for each channel realizations, we set $\tau_{arld}=\tilde{ARLD}$ in \eqref{attacksequential} to find the optimal $\bm{\Omega}$; this is done by computing $\sigma^2$ from Theorem~\ref{theorem:Attack Design against sequential_finding_bound_sigma2}, using it in \eqref{attacksequential_alternate_vectorform}, and running an algorithm similar to Algorithm~\ref{algorithm:For_finding_optimum_attack_phase}. We average the results over multiple independent channel realizations to find the expected percentage decrease of data rate under our attack scheme. Table~\ref{tab:Table_1_baseline_sequential} shows our attack results in significantly more percentage decrease in data rate compared to the baseline scheme, for the same values of ARLFA and expected ARLD. Moreover,  ARLD under random phase shift attack is very small and it cannot be increased as $\epsilon_{\mathrm{GLR}}$ is determined by ARLFA which is chosen by the receiver. However, optimal attack is provided by our algorithm here.

\begin{table}[ht!]
\Huge
\centering\caption{Decrease in ergodic data rate with CMA}
 \label{tab:Table_1_ergodic data rate decrease}
\resizebox{8.7cm}{!}{\begin{tabular}{|c|c|c|c|}
 \hline
 \begin{tabular}{@{}c@{}}$\zeta_l$ as\\percentage of\\$\bar{SNR}_l \hspace{1mm}\forall \hspace{1mm} l=1,2$\end{tabular}&\begin{tabular}{@{}c@{}}Ergodic data rate\\without attack\\ (bits/channel use)\end{tabular}&\begin{tabular}{@{}c@{}}Ergodic data rate\\with attack\\ (bits/channel use)\end{tabular} & \begin{tabular}{@{}c@{}} Decrease in\\ Ergodic data\\ rate\end{tabular} \\
 \hline
 \begin{tabular}{@{}c@{}}$\zeta_{1}=10\%$\\$\zeta_{2}=10\%$\\ \end{tabular} & 6.735&6.1591&9.30\%\\
 \hline
 \begin{tabular}{@{}c@{}}$\zeta_{1}=20\%$\\$\zeta_{2}=20\%$\\ \end{tabular} & 6.6781&5.6752&17.67\%\\
 \hline
 \begin{tabular}{@{}c@{}}$\zeta_{1}=30\%$\\$\zeta_{2}=30\%$\\ \end{tabular} & 6.6771&5.313&25.67\%\\
 \hline
 \begin{tabular}{@{}c@{}}$\zeta_{1}=40\%$\\$\zeta_{2}=40\%$\\ \end{tabular} & 6.5042&4.7832&35.98\%\\
 \hline
 \end{tabular}}
 \end{table}

\begin{table}[h!]
\Huge
\centering\caption{Comparison with random phase shift attack scheme for the scenario of Section~\ref{Attack at a large time window} }
\label{tab:Table_1_baseline_multiple_fading_blocks}
\resizebox{6.1cm}{!}{\begin{tabular}{|c|c|c|}
\hline
 \begin{tabular}{@{}c@{}} Values of \\ $\kappa_{l}$\\ \end{tabular} &
 \begin{tabular}{@{}c@{}}Percentage decrease \\in ergodic data rate \\for random phase \\shift attack model\\ \end{tabular}&\begin{tabular}{@{}c@{}}Percentage decrease \\in data rate \\for our algorithm \end{tabular}  \\
 \hline
 \begin{tabular}{@{}c@{}}$\kappa_{1}=0.56$\\$\kappa_{2}=0.52$\\ \end{tabular} &34.4\%&52.6\%\\
 \hline
 \begin{tabular}{@{}c@{}}$\kappa_{1}=0.61$\\$\kappa_{2}=0.58$\\ \end{tabular} &36.5\%&57.7\%\\
 \hline
 \end{tabular}}
\end{table}

\subsection{SNR based detector over multiple fading blocks} We consider an RIS with $N=8$ and four possible phase shifts ($b=2$). All results are averaged over 100 independent channel realizations. Table~\ref{tab:Table_1_ergodic data rate decrease} shows that less stringent constraints result in more reduction in the ergodic data rate. However, it is important to note that,  the number of variables in \eqref{attackoptimizationproblem} is $2^{bN}$. This requires development of low complexity algorithms for large $N$. We compare our algorithm with the   random phase shift attack   in Table~\ref{tab:Table_1_baseline_multiple_fading_blocks}. We choose the detection threshold at the receiver as $\bar{\zeta}_{l}=\bar{SNR}_{l}$, and compute the ergodic data rate and expected detection probability $\tilde{P}_D$ for random phase shift attack by averaging over   channel realizations. Next, we find $\zeta_{l}=\kappa_l \bar{\zeta}_l$ for  \eqref{attackoptimizationproblem} such that the expected probability of detection under our attack becomes exactly equal to $\tilde{P}_D$, and find the corresponding optimized  ergodic data rate.    Table~\ref{tab:Table_1_baseline_multiple_fading_blocks} shows  that  our attack scheme significantly outperforms   the baseline scheme.


\begin{table}[ht!]
\centering\caption{Decrease in data rate with CMA under CSI uncertainty }
 \label{tab:Table_1_data rate decrease KS Test}
\resizebox{8.5cm}{!}{\begin{tabular}{|c|c|c|c|c|c|c|c|}
 \hline
 \multicolumn{7}{|c|}{Percentage decrease in data rate for different values of ${\nu}_{ks}$ and $\bar{\varepsilon}_{ks}$} \\
 \hline
\begin{tabular}{@{}c@{}} \\$\hspace{0.1mm}$\\  \end{tabular}&  \begin{tabular}{@{}c@{}} ${\nu}_{ks}=0.10 $\end{tabular} & \begin{tabular}{@{}c@{}} ${\nu}_{ks}=0.15 $\end{tabular} & \begin{tabular}{@{}c@{}} ${\nu}_{ks}=0.20 $\end{tabular} & \begin{tabular}{@{}c@{}} ${\nu}_{ks}=0.25 $\end{tabular}& \begin{tabular}{@{}c@{}} ${\nu}_{ks}=0.30 $\end{tabular}& \begin{tabular}{@{}c@{}} ${\nu}_{ks}=0.35 $\end{tabular}\\
 \hline
 \begin{tabular}{@{}c@{}}\\ $\bar{\varepsilon}_{ks}=0.02$ \\$\hspace{0.5mm}$ \end{tabular} & 49.4386\%  & 44.840\%&41.578\%&39.046\%&36.985\%&35.2329\%\\
 \hline
 \begin{tabular}{@{}c@{}}\\$\bar{\varepsilon}_{ks}=0.04$\\$\hspace{0.5mm}$ \end{tabular} & 37.996\% & 33.350\%&30.054\%&27.505\%&25.410\%&23.644\%\\
 \hline
 \end{tabular}}
\end{table}

\subsection{Per sample double threshold test over a given fading block under imperfect CSI} 
We consider $N=64$  and  $K=100$,   simulate the attack designed in Section~\ref{CSI uncertainty}, and  determine its  performance  averaged over $10000$ independently generated channel realizations. For a significance level $\hat{\alpha}=0.01$ and corresponding critical value $\bar{\iota}=0.230$ for the K-S Test, we 
compute $r_{l}$ and $r_{u}$ for various values of $\bar{{\varepsilon}}_{ks}$. Finally, for various values of $\nu_{ks}$, we compute the optimum phase shift matrix ($\bm{\Omega}$) and employ it to attack the generated instances. Table~\ref{tab:Table_1_data rate decrease KS Test} shows us that, as    ${\nu}_{ks}$ increases, there is a decrease in the percentage reduction of data rate; this is due to the fact that as ${\nu}_{ks}$ increases the received samples  are of much higher energy and hence there is a smaller reduction in data rate. Also, we observe that as $\bar{{\varepsilon}}_{ks}$ increases, the percentage decrease in data rate reduces because the  detector becomes more cautious. Additionally, as   in Section~\ref{results_UMP}, here also  our algorithm performs   better in the low SNR regime. 
\begin{table}
\centering\caption{Comparison with random phase shift attack   for the scenario described in  Section~\ref{CSI uncertainty} }
\label{tab:Table_1_baseline_CSI_uncertainity}
\resizebox{8.4cm}{!}{\begin{tabular}{|c|c|c|c|}
 \hline
 \begin{tabular}{@{}c@{}} Value of \\$\bar{{\varepsilon}}_{ks}$\\ \end{tabular} &
 \begin{tabular}{@{}c@{}}Value of \\ ${\nu}_{ks}$ \end{tabular}&\begin{tabular}{@{}c@{}}Percentage decrease \\in data rate \\for random phase \\shift attack model\\ \end{tabular}&\begin{tabular}{@{}c@{}}Percentage decrease \\in data rate \\for our algorithm \end{tabular}  \\
 \hline
 \begin{tabular}{@{}c@{}}$\bar{{\varepsilon}}_{ks}=0.02$\\ \end{tabular} & 0.57&12.08\%&28.7\%\\
 \hline
 \begin{tabular}{@{}c@{}}$\bar{{\varepsilon}}_{ks}=0.04$\\ \end{tabular} & 0.52&8.86\%&17.8\%\\
 \hline
 \end{tabular}}
\end{table}
Table~\ref{tab:Table_1_baseline_CSI_uncertainity} shows a comparison between our algorithm and random phase shift attack scheme. For a particular value of $\bar{{\varepsilon}}_{ks}$ (which determines the false alarm rate) and channel realization, we compute the thresholds $r_{l}$ and $r_{u}$ by using Theorem~\ref{theorem:KS_Test_detector_energy_range}. Next, we simulate the value of ${\nu}_{ks}$ (which determines the detection probability, see Section~\ref{AttackDesignImperfect}) and mean data rate under random phase shift attack for a KS test based detector using the thresholds $r_{l}$ and $r_{u}$. By averaging over  multiple  channel realizations, we calculate the expected value of ${\nu}_{ks}$ ({\em i.e.,} $\hat{{\nu}}_{ks}$) and the expected percentage decrease in data rate. Next, for each channel realization,  we set ${\nu}_{ks}=\hat{{\nu}}_{ks} $ in \eqref{Unknown_CSI_opti} to find the optimal $\bm{\Omega}$. Finally, we solve the optimization problem \eqref{Unknown_CSI_opti}, using an algorithm similar to  Algorithm~\ref{algorithm:For_finding_optimum_attack_phase}. We average the results over multiple independent channel realizations to find the expected percentage decrease in data rate under our attack scheme.  Table~\ref{tab:Table_1_baseline_CSI_uncertainity} shows that our attack scheme significantly outperforms the baseline.

\section{Conclusion}\label{Conclusion}
In this paper, we have introduced    controller manipulation attack (CMA), and analytically derive detectors and attack design algorithms for CMA. We have considered various aspects of the system model: discrete versus continuous phase shift, single versus multiple transmit antennas, perfect versus imperfect CSI, one shot detection versus sequential detection of CMA, attack over single versus multiple fading blocks,  to name a few. While the structures of the proposed detectors  are simple, the  attacks against all these detectors over a single fading block have been shown to be obtained by solving the same optimization problem for which we have also proposed an algorithm. For detection over multiple fading blocks, the attack design problem turns out to be a linear program. 

While our paper lays the foundation to the theory for CMA attacks on RIS assisted communication, some more  possible   non-idealities  need to be studied  properly  before we use these algorithms in a real system. Some of the challenges include: (i) reduction of computational complexity of the attack design algorithm proposed against detection over multiple fading blocks, (ii) correlation across various SISO channels involved between Tx/Rx and the RIS, (iii) lack of knowledge of the fading block length and consequently the number of symbols transmitted over a fading block, (iv) lack of a precise statistical model for the CSI error,  (v) non-Rayleigh fading, possibly with a strong line-of-sight component as seen in millimeter wave communication,  (vi) near-field communication involving spherical wavefront model, and (vii) non-stationary channel statistics. We seek to address these practical issues in our future research.

\appendices

\section{Proof of Lemma~\ref{lemma:maxvariance}}
\label{appendix:proof-of-maxvariance}
We know, $\Lambda_k=\phi_{k}+\psi_{k}+\theta_{k}$. Now, 
$\sigma^2=\sigma_w^2+|\mathbf{{g}}^{H}{\mathbf{{}\Phi}\mathbf{h}}|^{2} = \sigma_w^2+|\sum_{k=1}^{N} \alpha_{k}\beta_{k} \mathrm{e}^{j \Lambda_{k}}|^{2}  \leq  \sigma_w^2+|\sum_{k=1}^{N} \alpha_{k} \beta_{k} \underbrace{|\mathrm{e}^{j \Lambda_{k}}|}_{=1} |^{2}$.\\
Obviously, $\sigma^2=\sigma_{0}^2$ only if $\Lambda_k=0$ for all $1 \leq k \leq N$, i.e.,  if $\bm{\Phi}=\bm{\Phi}_0$.

\section{Proof of lemma~\ref{lemma:likelihoodratio}}
\label{appendix:proof-of-likelihoodratio}

The likelihood ratio is:
    $L_{\sigma}(\bm{y})=\frac{p_{\sigma}(\bm{y};H_1)}{p_{\sigma_{0}}(\bm{y};H_0)}\nonumber=\frac{\frac{1} {\pi^K(\sigma^{2})^K}\hspace{0.1cm} \mathrm{e}^{-\frac{\sum_{i=1}^{K}||y_i||^2}{\sigma^2}}}{\frac{1} {\pi^K(\sigma_{0}^{2})^K}\hspace{0.1cm} \mathrm{e}^{-\frac{\sum_{i=1}^{K}||y_i||^2}{\sigma_{0}^2}}}$.
Taking logarithm and considering a generic threshold $\eta>0$,  the likelihood ratio test becomes:\\
    $\ln L_{\sigma}(\bm{y})=K\ln\frac{{\sigma_{0}}^2}{\sigma^2}+\sum_{i=1}^{K}||y_i||^2(\frac{1}{{\sigma_{0}}^2}-\frac{1}{\sigma^2}){\underset{H_{0}}{\overset{H_{1}}{\gtreqqless }}}\eta\nonumber \\
    \implies \sum_{i=1}^{K}||y_i||^2 {\underset{H_{0}}{\overset{H_{1}}{\lesseqqgtr}}} \underbrace{(\frac{1}{{\sigma_{0}}^2}-\frac{1}{\sigma^2})^{-1}(\eta-K\ln\frac{{\sigma_{0}}^2}{\sigma^2})}_{\doteq \eta'}\nonumber$


\section{Proof of Lemma~\ref{lemma:Yidistribuition}}
\label{appendix:proof-of-Yidistribuition}
Let $y_{i}\sim \mathcal{CN}(0,{\Tilde{\sigma}}^2)$ for any arbitrary ${\Tilde{\sigma}}^2 > 0$. Now, $ ||y_{i}||^2= {Y_{R}}^2 + {Y_{I}}^2$, where $Y_{R}$ and $Y_{I}$ are real and imaginary  parts of $y_{i}$. Obviously, $Y_{R}\sim \mathcal{N}(0,{\Tilde{\sigma}}^2/2)$, $Y_{I}\sim \mathcal{N}(0,{\Tilde{\sigma}}^2/2)$. Since ${Y_{R}}^2$ and  ${Y_{I}}^2$ are chi-squared distributed random variables with one degree of freedom, hence $||y_{i}||^2$ is an  exponentially distributed random variable with parameter $\lambda=\frac{1}{\Tilde{\sigma}^2}$ \cite[Chapter~6, Equation~6-68]{papoulis2002probability}. Now, $W \doteq \sum_{i=1}^{K}||y_i||^2$ is a sum of $K$ exponentially distributed random variables and as $K$ is an integer, hence the sum ($W$) generalizes to Erlang distribution with parameters as $K$ and $\lambda$\cite[Chapter~7, Equation~7-165]{papoulis2002probability}. We define the characteristic function (c.f.) $\varphi _{U}(\bar{r})$ of any random variable U as the Fourier transform of its probability density function $f_{U}(u)$, expressed as $\varphi _{U}(\bar{r})=\mathbb {E}\left[e^{i\bar{r}U}\right]=\int_{\mathbb{R}} e^{i \bar{r} u} f_{U}(u) du$. We can write the c.f. of $W$ as $\varphi _{W}(\bar{r})=\mathbb {E}\left[e^{i\bar{r}W}\right]=\tfrac{1}{(1-\frac{i\bar{r}}{\lambda})^K}$. Representing, $D=2\lambda W$ we can write its c.f. as: $\varphi _{D}(\bar{r})=\mathbb {E}\left[e^{i\bar{r}D}\right]=\mathbb {E}\left[e^{i\bar{r} 2 \lambda W}\right]$. Let $r'=2\lambda \bar{r}$, hence $\varphi_{D}(\bar{r})=\mathbb {E}\left[e^{ir'W}\right]=\tfrac{1}{(1-\frac{ir'}{\lambda})^K}=\tfrac{1}{(1-\frac{i 2\lambda \bar{r}}{\lambda})^K}=\tfrac{1}{(1-i 2 \bar{r})^K}$, which is the c.f. of a chi-squared random variable with $2K$ degrees of freedom \cite[Chapter~5, Table~5-2]{papoulis2002probability}. Hence, $2 \lambda W$ is a chi-squared random variable with $2K$ degrees of freedom.

\section{Proof of Theorem~\ref{theorem:UMP-Test}}
\label{appendix:proof-of-UMP-test}
From Lemma \ref{lemma:likelihoodratio}, we obtain 
    $P_{FA}=\mathbb{P}(W \leq \eta'| H_0) =\mathbb{P}(2\lambda_{0} W \leq\eta'' | H_0)=\rho
$, 
where $\lambda_{0}=1/{\sigma_{0}}^2$ and $\eta''=2 \lambda_0 \eta'$. Using this, the detector chooses the threshold $\eta'$ for a target  probability of false alarm $\rho$. We have $\eta''=R_{2K,\rho}$ and hence the threshold   $\eta'=\eta''/2\lambda_{0}$  is independent of   $\sigma^2$ (and  $\bm{\Phi}$) and  the  test is UMP.

\section{Proof of Theorem~\ref{theorem:Attack Design against UMP}}
\label{appendix:proof-of-Attack_Design_against_UMP}

We are given that $\xi=\mathbb{P}(2\lambda W \leq 2\lambda\eta'|H_1)$. Now, we know that  $2 \lambda\eta'={R_{2K,\xi}}$, where $\lambda=1/{\sigma}^2$ and  $\eta'=\frac{{R_{2K,\rho}}}{2\lambda_{0}}$. This yields: 
$\frac{R_{2K,\rho}}{2\lambda_{0}}=\frac{R_{2K,\xi}}{2\lambda} \hspace{2mm}\implies{{R_{2K,\rho}}{\sigma_{0}^2}}={{R_{2K,\xi}}{\sigma^2}}\hspace{2mm}\implies{\sigma^2}= \frac{{R_{2K,\rho}}{\sigma_{0}^2}}{{R_{2K,\xi}}}$

\section{Proof of Theorem~\ref{theorem:Attack Design against sequential_finding_a}}
\label{appendix:proof-of-Attack_Design_against_sequential_finding_a}

From \cite[Theorem~1]{lorden1973open}, we have $b_{\mathrm{GLR}}=3 \ln ({a_{\mathrm{GLR}}}^{-1})\left(1+\frac{1}{I({\sigma_{min}^2})} \right)^{2}$ and   $\epsilon_{\mathrm{GLR}}=-\ln \{a_{\mathrm{GLR}}/b_{\mathrm{GLR}}\}$. By putting these values and performing some simple algebra, we obtain:  $
    \ln {b_{\mathrm{GLR}}} - \ln {a_{\mathrm{GLR}}} = \epsilon_{\mathrm{GLR}}  
    \implies \frac{\ln (a_{\mathrm{GLR}})}{a_{\mathrm{GLR}}}=-\frac{e^{\epsilon_{\mathrm{GLR}}}}{3\big(1+\frac{1}{I({\sigma_{min}^2})}\big)^{2}}\hspace{20mm} $

\section{Proof of Theorem~\ref{theorem:Attack Design against sequential_finding_bound_sigma2}}
\label{appendix:proof-of-Attack_Design_against_sequential_finding_bound_sigma2}
We know from \cite[Theorem~1]{lorden1971procedures} that the lower bound on ARLD is given by $\mathbb{E}_{f_{1}}(T_{\mathrm{GLR}}) \geq \frac{-\ln {a_{\mathrm{GLR}}}}{I(\sigma^2)}$. Hence, for $ARLD \geq \tau_{arld}$, it is sufficient to ensure $\frac{-\ln {a_{\mathrm{GLR}}}}{I(\sigma^2)} \geq \tau_{arld}$, i.e., $I(\sigma^2)\leq -\frac{\ln {a_{\mathrm{GLR}}}}{\tau_{arld}}$. 
 Now, $I(\sigma^2)$ decreases as   $\sigma^2$ increases towards $\sigma_0^2$,  and    the parameter \enquote{$a_{\mathrm{GLR}}$} is chosen using Theorem ~\ref{theorem:Attack Design against sequential_finding_a}. Hence, by using the expression of $I(\sigma^2)$, we can write:
$
    \ln\big(\frac{{\sigma_{0}}^2}{{\sigma}^2}\big)+\big(\frac{{\sigma}^2-{\sigma_{0}}^2}{{\sigma_{0}}^2}\big) \leq -\frac{\ln {a_{\mathrm{GLR}}}}{\tau_{arld}}  
    \implies {{\sigma}^2}-{{\sigma_{0}}^2}\ln ({{\sigma}^2}) \leq    \underbrace{ {{\sigma_{0}}^2}(1-\ln{{\sigma_{0}}^2}-\frac{\ln {a_{\mathrm{GLR}}}}{\tau_{arld}}) }_{\doteq L'}.$
This reduces to the condition $\sigma^2 \geq Q'$ for a suitable threshold $Q'>0$.

\section{Proof of Theorem~\ref{theorem:continuous}}
\label{appendix:proof-of-continuoussnr}
Note that, $\alpha_{k}$ and $\beta_{k}$ are  independent Rayleigh distributed random variables and ${X}_{k}=\alpha_{k}\beta_{k}$. Also,   $\mathbb{E}({X}_{k})=\mu_{x}=\sqrt{\epsilon_{h}\epsilon_{g}}\frac{\pi}{4}$ and   $Var({X}_{k})={\sigma_{x}}^2=\epsilon_{h}\epsilon_{g}(1-\frac{{\pi}^2}{16})$ \cite{salo2006distribution}. By applying Central Limit Theorem:
$\Xi_N \doteq \frac{Z-N\mu_{x}}{\sqrt{N}\sigma_{x}}\overset{d}{\longrightarrow}  \mathcal{N}(0,1)$.
Hence, for large $N$, we can write the mean and variance of $Z$ as $\mu_{z}=N \sqrt{\epsilon_{h}\epsilon_{g}}\frac{\pi}{4}$ and $\sigma_{z}^2=N\epsilon_{h}\epsilon_{g}(1-\frac{{\pi}^2}{16})$. Also, since SNR ($\Gamma^{*}$)=$\bar{\kappa}|\sum_{k=1}^{N}\alpha_{k}\beta_{k}|^{2} = \bar{\kappa}|Z|^{2}$, the SNR follows a non-central chi-squared distribution with one degree of freedom.
Hence, we can write its P.D.F. as $
    f_{\Gamma^{*}}(\gamma)=\frac{1}{2\sqrt{2\pi\bar{\kappa}\sigma_{z}^2\gamma}}\bigg({e^{-\frac{{(\sqrt{\frac{\gamma}{\bar{\kappa}}}-\mu_{z}})^2}{2\sigma_{z}^2} }}+{e^{-\frac{{(\sqrt{\frac{\gamma}{\bar{\kappa}}}+\mu_{z}})^2}{2\sigma_{z}^2}}}\bigg)$ 
and its M.G.F as $M_{\Gamma^{*}}(t)=e^{\mu_Z^2\bar{\kappa} t /(1-2\bar{\kappa} t\sigma_Z^2)}(1-2\bar{\kappa} t\sigma_Z^2)^{-\frac{1}{2}}$. 
Hence, the first moment $\bar{SNR}_1$ and second moment $\bar{SNR}_2$ of SNR can be expressed as:
\begin{equation*}
\mathbb{E}(\Gamma^{*})=\bar{\kappa}\mathbb{E}(Z^2)
=\bar{\kappa}(\mu_{z}^2+\sigma_{z}^2)
=N \bar{\kappa} \epsilon_{h}\epsilon_{g}(1+\frac{\pi^2}{16}(N-1))
\end{equation*}
\begin{align*}
    \mathbb{E}({\Gamma^{*}}^2)=&N^2\bar{\kappa}^2\epsilon_{h}^2\epsilon_{g}^2\bigg(1+\frac{\pi^4}{256}(N-1)^2 + \frac{\pi^2}{8}(N-1)+\nonumber\\
    &\frac{16-\pi^2}{8}+\frac{\pi^2(2N-1)(16-\pi^2)}{128}\bigg)\nonumber
\end{align*}

\section{Proof of Theorem~\ref{theorem:discrete}}
\label{appendix:proof-of-discretesnr}
Let us define $V \doteq \frac{1}{N}\sum_{k=1}^{N}V_k$, where $V_k={\alpha_k}{\beta_k}e^{j\delta_{k}}$. We observe that as the distribution of $\delta_k$ is symmetric around the origin, and hence its characteristic function $\varphi_{\delta_k}(\omega)$ is always real. Hence,  $\mu=\mathbb{E}(V_k)=\mathbb{E}({\alpha_k})\mathbb{E}({\beta_k})\mathbb{E}(e^{j\delta_k})=\frac{\pi}{4}\sqrt{\epsilon_{h}\epsilon_{g}}{\varphi_{\delta_k}(1)}$. Similarly, variance $\upsilon={\epsilon_h}{\epsilon_g}(1-\frac{\pi^2}{16}|{\varphi_{\delta_k}(1)}|^2)$ and pseudo-variance $\hat{\rho}={\epsilon_h}{\epsilon_g}({\varphi_{\delta_k}(2)}-\frac{\pi^2}{16}{\varphi_{\delta_k}(1)}^2)$ \cite[Chapter~7, Page~249]{papoulis2002probability}. Obviously,  $\mu,\upsilon,\hat{\rho} \in \mathbb{R}$. By applying central limit theorem, for large $N$ we can approximate the distribution of $V$ as  $\mathcal{CN}(\mu,\frac{\upsilon}{N},\frac{\hat{\rho}}{N})$. Let $V_R$ and $V_I$ denote the real and imaginary parts of $V$.  We can write:
\begin{equation}
\mathrm{Cov}(V_R,V_I)=\frac{1}{2}\mathfrak{Im}(-\frac{\upsilon}{N}+\frac{\hat{\rho}}{N})=0.
\end{equation}
Since $V_R$ and $V_I$ are jointly Gaussian and uncorrelated, they are mutually  independent. Also, $V_R \sim \mathcal{N}(\mu_{V_R},\sigma_{V_R}^2)$ and $V_I \sim \mathcal{N}(0,\sigma_{V_I}^2)$, where $\mu_{V_R}=\mu$ and 
\begin{equation}
\sigma_{V_R}^2=\frac{1}{2}\mathfrak{Re}(\frac{\upsilon}{N}+\frac{\hat{\rho}}{N})=\frac{{\epsilon_h}{\epsilon_g}}{2N}(1+{\varphi_{\delta_k}(2)}-\frac{\pi^2}{8}{\varphi_{\delta_k}(1)}^2)
\end{equation}
\begin{equation}
\sigma_{V_I}^2=\frac{1}{2}\mathfrak{Re}(\frac{\upsilon}{N}-\frac{\hat{\rho}}{N})=\frac{{\epsilon_h}{\epsilon_g}}{2N}(1-{\varphi_{\delta_k}(2)})
\vspace{+2mm}
\end{equation}
By using the results derived in
\cite[Theorem~1, Appendix~B]{badiu2019communication} and \cite{moschopoulos1985distribution} we can approximate the distribution of $|V|^{2}$ (for large $N$) as a gamma distribution with shape parameter  $k_{{V}^2}=\frac{\mu_{V_R}^{2}}{4 \sigma_{V_R}^{2}}$ and scale parameter as $\theta_{{V}^2}=4 \sigma_{V_R}^{2}$. From \eqref{discretereceivedSNR} we write the SNR as $\Gamma^{*}=\bar{\kappa}{N}^2|V|^{2}$.  After rewriting, we obtain  $\Gamma^{*}=\varrho |V|^2$, where $\varrho=\bar{\kappa}{N}^2$ and $\varrho >0$. Hence, we can infer from above that as $|V|^2$ follows gamma distribution hence the SNR also follows a gamma distribution with shape parameter as $k_{\Gamma^{*}}=k_{{V}^2}$ and scale parameter as $\theta_{\Gamma^{*}}=\varrho\theta_{{V}^2}$. As a result, we can write the PDF of SNR as:
\begin{equation}
    f_{\Gamma^{*}}(\gamma)={\frac {\gamma^{{k_{\Gamma^{*}}}-1}e^{-\frac{\gamma}{\theta_{\Gamma^{*}}} }}{\theta_{\Gamma^{*}} ^{k_{\Gamma^{*}}}\Gamma (k_{\Gamma^{*}})}}
    ={\frac {\gamma^{{k_{V^{2}}}-1}e^{-\frac{\gamma}{\bar{\kappa}{N}^2\theta_{V^{2}}}}}{{(\bar{\kappa}{N}^2\theta_{V^{2}})} ^{k_{V^{2}}}\Gamma (k_{V^{2}})}}\nonumber
\end{equation}
Hence, the M.G.F of \hspace{1.5mm} $|V|^2$ is \hspace{1.5mm}  $M_{|V|^{2}}(t)=(1-t \theta_{|V|^{2}})^{-k_{|V|^{2}}},\hspace{1.5mm}~\forall t\hspace{1.5mm}<{\tfrac {1}{\theta_{|V|^{2}}}}$ where \\ $\tfrac {1}{\theta_{|V|^{2}}}=\tfrac{N}{{\epsilon_h}{\epsilon_g}(1+{\varphi_{\delta_k}(2)}-\frac{\pi^2}{8}{\varphi_{\delta_k}(1)}^2)}$ is large for large $N$. Further, the M.G.F of SNR $\Gamma^{*}$ can be expressed as $M_{\Gamma^{*}}(t)=(1-t\varrho \theta_{|V|^{2}})^{-k_{|V|^{2}}}$.
\cite[Chapter~5, Table~5-2]{papoulis2002probability}.

The $l^{\text{th}}$ moment of SNR can be expressed as:$\bar{SNR}_{l}=\mathbb{E}\left({\Gamma^{*}}^{l}\right)=M_{\Gamma^{*}}^{(l)}(0)=\left.\frac{d^{l} M_{\Gamma^{*}}}{d t^{l}}\right|_{t=0}$.
Hence, the first moment $\bar{SNR}_{1}$ and second moment $\bar{SNR}_{2}$ are: 
\begin{align}
    \mathbb{E}(\Gamma^{*})=&\varrho k_{{V}^2} \theta_{{V}^2} 
    =\bar{\kappa}{N}^2 \mu_{V_R}^{2}
    =\frac{N^2\bar{\kappa} \pi^2}{16}\epsilon_h\epsilon_g{\varphi_{\delta_k}(1)}^2\nonumber\\
    \mathbb{E}({\Gamma^{*}}^2)
    =&\varrho^2 k_{{V}^2} \theta_{{V}^2}^2\big(1+k_{{V}^2}\big) 
    =\bar{\kappa}^2 {N}^4 \mu_{V_R}^2\big(4\sigma_{V_R}^{2}+\mu_{V_R}^2\big)\nonumber\\
    =&\epsilon_h^2\epsilon_g^2\bigg(\frac{N^3 \pi^2}{4}\bigg(1+{\varphi_{\delta_k}(2)}-\frac{\pi^2 {\varphi_{\delta_k}(1)}^2}{8}\bigg)+ \nonumber\\
    &\frac{N^4\pi^4{\varphi_{\delta_k}(1)}^4}{256}\bigg)\nonumber
\end{align}

\section{Proof of Theorem~\ref{theorem:pdf_under_CSIR_uncertainty}}\label{appendix:proof-of-pdf_CSIR_uncertainty}\label{pdf_proof_csi_uncertainty}
From   \eqref{Received_signal_CSI_uncertainty}, we can write, $y_{uc}=\bm{s}^{\mathrm{H}} \hat{\bm{\Psi}}x+ \bm{s}^{\mathrm{H}}\Delta\bm{\Psi}x+w = \Xi_1 + \Xi_2 + w $
where $\Xi_1=\bm{s}^{\mathrm{H}} \hat{\bm{\Psi}}x$ and $\Xi_2=\bm{s}^{\mathrm{H}}\Delta\bm{\Psi}x$. Now, $\Xi_1=\bm{s}^{\mathrm{H}} \hat{\bm{\Psi}}x=(\sum_{i=0}^{N} \hat{\Psi_i}\mathrm{e}^{j \phi_{i}})x$, where $|\sum_{i=1}^{N} \hat{\Psi_i}\mathrm{e}^{j \phi_{i}}|^2$ is a constant and $x \sim \mathcal{N}(0,1)$. Hence,  $\Xi_1\sim \mathcal{CN}(0,|\sum_{i=1}^{N} \hat{\Psi_i}\mathrm{e}^{j \phi_{i}}|^2)$. 

Similarly, we have $\Xi_2=\bm{s}^{\mathrm{H}}\Delta\bm{\Psi}x$. Let $\bar{Z}=\bm{s}^{\mathrm{H}}\Delta\bm{\Psi}=\sum_{i=0}^{N} {\Delta\Psi_i}\mathrm{e}^{j \phi_{i}}$, where $\Delta\bm{\Psi}\sim \mathcal{C} \mathcal{N}(\bm{0}, {\sigma_{e}^2}\bm{I})$. Hence,  
   $\operatorname{Var}(\bar{Z})=\operatorname{Var}[\sum_{i=1}^{N} \mathrm{e}^{j \phi_{i}}  \Delta\Psi_i]= \sum_{i=1}^{N} \sum_{k=1}^{N} \mathrm{e}^{j \phi_{i}} {\mathrm{e}^{-j \phi_{k}}} \operatorname{Cov}[\Delta\Psi_i, \Delta\Psi_k]$ and $\mathbb{E}(\bar{Z})=\sum_{i=1}^{N} \mathbb{E}({\Delta\Psi_i})\mathbb{E}(\mathrm{e}^{j \phi_{i}})=0$. As, $\operatorname{Cov}[\Delta\Psi_i, \Delta\Psi_k]=\sigma_e^2 \mathbbm{1}_{\{i=k\}}$, we can write $\operatorname{Var}(\bar{Z})=N\sigma_e^2$ and  $ \bar{Z}\sim \mathcal{CN}({0},N\sigma_e^2)$. 

The received signal is a complex random variable expressed as follows:
\footnotesize
\begin{align}\label{Received_signal_y_unknown_CSI}
y_{uc}&=\underbrace{\Xi_1}_{\sim \mathcal{CN}(0,|\sum_{i=1}^{N} \hat{\Psi_i}\mathrm{e}^{j \phi_{k}}|^2)} + \underbrace{\bar{Z}}_{\sim \mathcal{CN}(0,N\sigma_e^2)}\underbrace{x}_{\sim \mathcal{CN}(0,1)} + \underbrace{w}_{\sim \mathcal{CN}(0,\sigma_w^2)}\nonumber\\
&=\underbrace{\Xi_1+w}_{\sim \mathcal{CN}(0,\sigma_w^2+|\sum_{i=1}^{N} \hat{\Psi_i}\mathrm{e}^{j \phi_{k}}|^2)}+\underbrace{\bar{Z}}_{\sim \mathcal{CN}(0,N\sigma_e^2)}\underbrace{x}_{\sim \mathcal{CN}(0,1)}
\end{align}
\normalsize

We denote $\sigma^{2}=\sigma_{w}^{2} + |\sum_{i=0}^{N}\hat{\psi_{i}}\mathrm{e}^{j \phi_{k}}|^2$ and $\sigma_{s}^{2}=N{\sigma_{e}}^2$. The closed form expression of the P.D.F. of $y_{uc}$ is obtained from \cite[Equation~15]{shi2020distribution} and can be written as:
\begin{align*}
f_{y_{uc}}(y_{uc})&=\frac{1}{\pi\sigma_{s}^2} \exp (\frac{\sigma^{2}}{\sigma_{s}^2}) \sum_{i=0}^{\infty}(\frac{1}{\sigma_{s}^{2} })^{i} G(-i, \frac{\sigma^{2}}{\sigma_{s}^{2}}) \frac{(-{y_{uc}}^{2})^{i}}{i!}
\end{align*}
where $G(\cdot, \cdot)$ denotes the incomplete Gamma function \cite{harris2008incomplete}. Similarly, the cumulative distribution function of energy of the received signal is obtained by \cite[Equation~19]{shi2020distribution} and is expressed as:
\begin{align*}
F_{{r,\sigma}}(r)&=1-[\frac{\sigma^{2}}{\sigma_{s}^{2}}\mathrm{e}^{-(\frac{\sigma^{2}}{\sigma_{s}^{2}})} \int_{1}^{\infty}(\mathrm{e}^{(-\frac{\sigma^{2}t}{\sigma_{s}^{2}}-\frac{r}{\sigma^{2}t})})\,dt   ]
\end{align*}
Under no attack,   RIS knows   $\hat{\bm{\Psi}}$ and  sets its phase shift in accordance to $\hat{\bm{\Psi}}$, namely $s_{i}=\mathrm{e}^{-j \arg(\hat{\psi_{i}})}, \, \forall \,\, 1\leq i \leq N$. Hence, the value of $\sigma^{2}$ reduces to $\sigma_{0}^{2}=\sigma_{w}^{2} + |\sum_{i=0}^{N}|\hat{\psi_{i}}||^2$.

\section{Proof of Theorem~\ref{theorem:KS_Test_detector_energy_range}}\label{appendix:KS_Test_detector_energy_range}

For a given $r_n$, $H_0$ and large enough $K$, the empirical CDF
$\hat{F}_{1}(\cdot)$ computed using all samples other than $r_n$ yields an unbiased estimate of $F_{r,\sigma_0}(\cdot)$. Hence,   given $r_n$ and $H_0$, ${\hat{D}_n}$ can be viewed as a zero mean random variable.
Thus, by Chebyshev's inequality, 
\begin{align}\label{chebyshev}
    \mathbb{P}(|\hat{D}_n| \geq \bar{\iota} \hspace{+2mm} |\hspace{+2mm} r_n,\hspace{+1mm} H_0) \leq \hspace{+2mm} \frac{\operatorname{Var}(\hat{D}_n \hspace{+1mm} |\hspace{+1mm} r_n,\hspace{+1mm} H_0)}{\bar{\iota}^2}
\end{align}
Given $r_n$ and $H_0$, $F_{r,\sigma_0}(r_n)$ is a constant. Hence, $\operatorname{Var}(\hat{D}_n \hspace{+1mm} |\hspace{+1mm} r_n,\hspace{+1mm} H_0) $ $= \operatorname{Var}(\hat{F}_1(r_n)|r_n, H_0) $ $ = \operatorname{Var}(\frac{1}{K-1}$ $ \sum_{j=1,j \neq n}^{K} \mathbb{I}\left(r_{j} \leq r_n\right)|r_n, H_0)=\frac{1}{K-1}\operatorname{Var}( \mathbb{I}\left(r_{1} \leq r_n\right)|r_n, H_0)=\frac{1}{K-1}(F_{r,\sigma_{0}}(r_n)-(F_{r,\sigma_{0}}(r_n))^2)$. The last equality follows from the fact that given $r_n$ and $ H_0$, the random variable $\mathbb{I}\left(r_{1} \leq r_n\right)$ follows a Bernoulli distribution with mean $F_{r,\sigma_{0}}(r_n)$. Therefore,
\begin{align}\label{chebyshev1}
    \mathbb{P}(|\hat{D}_n| \geq \bar{\iota} \hspace{+2mm} |\hspace{+2mm} r_n, H_0) \leq \hspace{+2mm} \frac{F_{r,\sigma_0}(r_n)-(F_{r,\sigma_0}(r_n))^2}{{(K-1)\bar{\iota}^2}}
\end{align}
Now, a sufficient condition to ensure $\mathbb{P}(|\hat{D}_n| \geq \bar{\iota}  |\  r_n, H_0) \leq \bar{\varepsilon}_{ks} $ is:
\begin{align}
    \frac{F_{r,\sigma_0}(r_n)-(F_{r,\sigma_0}(r_n))^2}{{(K-1)\bar{\iota}^2}} &\leq \bar{\varepsilon}_{ks}\nonumber\\
     \implies (F_{r,\sigma_0}(r_n))^2-F_{r,\sigma_0}(r_n) + &(K-1)\bar{\varepsilon}_{ks} {{\bar{\iota}^2}} \geq 0 \nonumber
\end{align}
Substituting $F_{r,\sigma_0}(r_n)=z_{n}$, we obtain:
\begin{equation}\label{quadratic}
\vspace{-2mm}
    {z_{n}}^2-z_{n}+(K-1)\bar{\varepsilon}_{ks} {{\bar{\iota}^2}} \geq 0
\end{equation}
We observe that \eqref{quadratic} is a quadratic inequality, which can be solved to obtain the range of feasible $z_{n}$ values as either $z_{n} \geq \frac{1+\sqrt{1-4(K-1)\bar{\varepsilon}_{ks} {{\bar{\iota}^2}}}}{2}$ or $z_{n} \leq \frac{1-\sqrt{1-4(K-1)\bar{\varepsilon}_{ks} {{\bar{\iota}^2}}}}{2}$. 
Using the fact that ${F}_{r,\sigma_0}(\cdot)$ is invertible because it  is a strictly increasing function, we obtain the upper threshold as $r_{u}=F^{-1}_{r,\sigma_0}\bigg(\frac{1+\sqrt{1-4(K-1)\bar{\varepsilon}_{ks} {{\bar{\iota}^2}}}}{2}\bigg)$ and the lower threshold as $r_{l}=F^{-1}_{r,\sigma_0}\bigg(\frac{1-\sqrt{1-4(K-1)\bar{\varepsilon}_{ks} {{\bar{\iota}^2}}}}{2}\bigg)$.

\ifCLASSOPTIONcaptionsoff
  \newpage
\fi



%


\bibliographystyle{IEEEtran}
\bibliography{ref.bib}


\end{document}